\documentclass[aps,amsmath,amssymb,superscriptaddress,floatfix, prb,10pt,twocolumn,eqrefnum]{revtex4-2}
\usepackage{amsmath}%
\usepackage{amsthm}%
\usepackage{xcolor}%
\usepackage{hyperref}
\usepackage{graphicx} 
\usepackage{bm}
\usepackage{physics}
\usepackage{bbm}
\usepackage{color}
\usepackage{comment}
\renewcommand{\vec}[1]{\textbf{#1}}
\newcommand{\vp}{\textbf{p}}
\newcommand{\hvp}{\hat{\textbf{p}}}

\renewcommand{\S}{\mathrm{S}}
\newcommand{\ch}{\mathrm{ch}}
\renewcommand{\sp}{\mathrm{sp}}
\newcommand{\AM}{\mathrm{AM}}

\begin{document}

\title{Spin-polarized supercurrents and Josephson diode effect in altermagnets
}
\author{Janus F. Niebuhr}
\affiliation{Institut f\"ur Physik, Universit\"at Greifswald, Felix-Hausdorff-Straße 6, 17489 Greifswald, Germany}
\author{Matthias Eschrig}
\affiliation{Institut f\"ur Physik, Universit\"at Greifswald, Felix-Hausdorff-Straße 6, 17489 Greifswald, Germany}
\author{Danilo Nikoli\'c}
\affiliation{Institut f\"ur Physik, Universit\"at Greifswald, Felix-Hausdorff-Straße 6, 17489 Greifswald, Germany}
\date{\today}
\begin{abstract}
We present a systematic theoretical study of the Josephson effect in junctions consisting of a d-wave altermagnet (AM) placed between two BCS superconductors (SC). In general, the SC/AM interfaces are spin-active and modeled by spin-dependent $\delta$ potentials, allowing for an arbitrary direction of the local exchange field vector. The model is formulated within the fully quantum (Gor'kov) and quasiclassical (Eilenberger) Green's function technique, applied to two distinct cases of (i) a weakly spin-polarized AM (exchange field much smaller compared to the Fermi energy) and (ii) a strongly spin-polarized AM (exchange field comparable to the Fermi energy). We apply our model to the SC/AM/SC geometry, accounting for the Josephson current-phase relation (CPR). In the weakly spin-polarized regime, the CPR displays the normal Josephson effect. Irrespective of the orientation of the altermagnet, the junction undergoes the $0-\pi$ transition. Depending on the orientation, the system displays the features similar to those of a ferromagnetic or an antiferromagnetic junction. To investigate the spin-polarized currents and nonreciprocal transport as the central results of the present work, we put the main focus on the strongly spin-polarized regime. Within this regime, we distinguish two cases. A coplanar exchange field profile across the junctions displays the normal Josephson effect; however, with a pure and stable long-range second harmonic in the CPR. In contrast, a noncoplanar exchange field profile gives rise to the so-called quantum geometric phases across the junction, leading to the absence of the phase-inversion center in the Josephson CPR. As a result, a Josephson diode effect emerges with a significant charge diode efficiency larger than $30\%$ and a perfect spin diode efficiency of $100\%$.
\end{abstract}

\maketitle

\section{Introduction}
The creation and control of long-range spin-triplet currents play the essential role both in the fundamental understanding and in the application of superconducting spintronics~\cite{eschrigSingletTripletMixingSuperconductor2004,buzdinProximityEffectsSuperconductorferromagnet2005,bergeretOddTripletSuperconductivity2005,eschrigSpinpolarizedSupercurrentsSpintronics2011,eschrigSpinpolarizedSupercurrentsSpintronics2015,linderSuperconductingSpintronics2015,birgereview2018,linderOddfrequencySuperconductivity2019,yangBoostingSpintronicsSuperconductivity2021,caiSuperconductorFerromagnetHeterostructures2023}. One of the most prominent playgrounds for studying such currents are superconductor (SC)-ferromagnet (FM) proximity contacts with a conventional superconducting order parameter, where the pair amplitudes are classified into singlet and triplet spin correlations~{\cite{tokuyasuProximityEffectFerromagnetic1988,demlerSuperconductingProximityEffects1997,bergeretLongRangeProximityEffects2001,eschrigSymmetriesPairingCorrelations2007,tanakaTheoryProximityEffect2007}}. Namely, the proximity effect allows for the conversion of spin-singlet correlations from the SC into triplet pair correlations in the FM. Considering the ferromagnetic exchange field in the FM as the spin quantization axis, one classifies these triplet correlations into short-ranged ($\uparrow\downarrow+\downarrow\uparrow$, $s_z=0$) and long-ranged ($\uparrow\uparrow$ and $\downarrow\downarrow$, $s_z=\pm 1$)~\cite{bergeretLongRangeProximityEffects2001}. The creation of long-range equal-spin triplets is allowed due to two fundamental processes occurring at the SC/FM interface~\cite{eschrigSpinpolarizedSupercurrentsSpintronics2011}: (i) the spin-mixing (or spin-dependent-phase-shift) effect due to the spin polarization of the SC/FM interface that creates the mixed-spin triplet correlations with $s_z=0$, and (ii) the spin-rotation mechanism that converts these correlations into the equal-spin ones with $s_z=\pm 1$. Note that the latter is possible only if a magnetic inhomogeneity is present. These two mechanisms have now been well established and experimentally verified~\cite{keizerSpinTripletSupercurrent2006,khaireObservationSpinTripletSuperconductivity2010,anwarLongrangeSupercurrentsHalfmetallic2010,robinsonControlledInjectionSpinTriplet2010,Glick2018,Caruso2019,Aguilar2020}.

Recently, considerable attention of the community was attracted to the discovery of a novel unconventional magnetic ordering termed altermagnetism~\cite{smejkalConventionalFerromagnetismAntiferromagnetism2022,smejkalEmergingResearchLandscape2022,fengAnomalousHallEffect2022}. This novel collinear magnetic phase derives from the effective decoupling of spin and orbital degrees of freedom due to negligible spin-orbit coupling, requiring a symmetry description in terms of the spin groups~\cite{brinkmanSpaceGroupTheory1966,brinkmanTheorySpinSpaceGroups1966,litvinSpinGroups1974,litvinSpinPointGroups1977,liuSpinGroupSymmetryMagnetic2022}. This decoupling allows for the classification of electronic bands in collinear magnetic arrangements into three types, (collinear) ferromagnets (FM), antiferromagnets (AFM) and altermagnets (AM), where the latter allows for spin-split but symmetry-compensated bands~\cite{smejkalConventionalFerromagnetismAntiferromagnetism2022,smejkalEmergingResearchLandscape2022}. This opens up the possibility of anomalous Hall effects with no net-magnetization~\cite{Smejkal2020,fengAnomalousHallEffect2022}, spin transfer torque effects~\cite{baiObservationSpinSplitting2022}, efficient spin-to-charge conversion~\cite{baiEfficientSpintoChargeConversion2023}, and merits of antiferromagnetic-like robustness against perturbations, no stray fields and fast dynamics for spintronics~\cite{baltzAntiferromagneticSpintronics2018}.

Considering superconducting junctions in general, if the system is invariant under time reversal and inversion, the Josephson current-phase relation (CPR) possesses a phase-inversion center at zero superconducting phase difference, $\Delta\chi=0$, i.e., $I(-\Delta\chi)=-I(\Delta\chi)\implies$ $I(\Delta\chi=0)=0$, and this is known as the normal Josephson effect~\cite{golubovCurrentphaseRelationJosephson2004}. However, if the mentioned symmetries are broken, the phase inversion center can be shifted to a different value of the superconducting phase difference, $\chi_0$, leading to $I(-\Delta\chi)\neq I(\Delta\chi) \implies I(\Delta\chi=0)\neq 0$, and this effect is known as the anomalous Josephson effect or the $\chi_0$ junction~\cite{GeshkenbeinLarkin1986,Yip1995,Sigrist1998,Buzdin2008}. If the Josephson CPR possesses no phase-inversion center, the critical current in one direction $(+)$ may differ from that in the opposite direction $(-)$~ leading to the so-called Josephson diode effect~\cite{nadeemSuperconductingDiodeEffect2023,shafferTheoriesSuperconductingDiode2025}. Recently, this effect has attracted considerable attention as the subject of intense experimental~\cite{andoObservationSuperconductingDiode2020, baumgartnerSupercurrentRectificationMagnetochiral2022, costaSignReversalJosephson2023, gutfreundDirectObservationSuperconducting2023, houUbiquitousSuperconductingDiode2023, nadeemSuperconductingDiodeEffect2023, strambiniSuperconductingSpintronicTunnel2022, trahmsDiodeEffectJosephson2023,reinhardtLinkSupercurrentDiode2024}  and theoretical investigations~\cite{greinSpinDependentCooperPair2009,costaSignReversalJosephson2023, daidoIntrinsicSuperconductingDiode2022, fominovAsymmetricHigherharmonicSQUID2022, haltermanSupercurrentDiodeEffect2022, hePhenomenologicalTheorySuperconductor2022, ilicTheorySupercurrentDiode2022, karabassovHybridHelicalState2022, kopasovGeometryControlledSuperconducting2021, misakiTheoryNonreciprocalJosephson2021, tanakaTheoryGiantDiode2022, yuanSupercurrentDiodeEffect2022, zhangGeneralTheoryJosephson2022, zinklSymmetryConditionsSuperconducting2022,soutoJosephsonDiodeEffect2022,steinerDiodeEffectsCurrentBiased2023,costaMicroscopicStudyJosephson2023,kopasovAdiabaticPhasePumping2023,Meyer2024,putilovNonreciprocalElectronTransport2024,tjernshaugenSuperconductingPhaseDiagram2024,patil2024,schulz2025_prl,schulz2025_prb}. 

 Considering superconducting hybrid structures involving unconventional magnets, various aspects of the proximity effect and the Josephson effect were investigated. The superconducting proximity effect was studied in Refs.~\cite{Chourasia2025,Sun2023,Papaj2023,zhangFinitemomentumCooperPairing2024,Vasiakin2025,Sukhachov2024,AlamProximitySuperconductivityInAltermagnets2026}. At the same time, the symmetry consideration of the pairing correlations was reported in Ref.~\cite{zhangFinitemomentumCooperPairing2024,maedaClassificationPairSymmetries2025,chakrabortyConstraintsSuperconductingPairing2025,fukayaSuperconductingPhenomenaSystems2025b,AlamProximitySuperconductivityInAltermagnets2026}. The normal Josephson effect has been predicted in different setups as shown in Refs.~\cite{Ouassou2023,SunSecondharmonic_2025,Cheng2024,liSpinPolarizedJosephsonSupercurrent2026}. In contrast, the anomalous Josephson effect ("$\phi_0$-junction") was reported in Refs.~\cite{Lu2024,alipourzadehAndreevBoundStates2025,fukayaJosephsonEffectOddfrequency2025}. The Josephson diode effect has also been studied in various systems based on altermagnets and other unconventional magnetic orderings, as reported in Refs.~\cite{sharmaTunableJosephsonDiode2025,boruahFieldFreeJosephson2025,chakrabortyPerfectSuperconductingDiode2024,banerjeeAltermagneticSuperconductingDiode2024,jiangJosephsonDiodeEffect2025a,debnathSpinpolarizationDiodeEffect2025,sahooFieldfreeTransverseJosephson2025,sharmaPwaveMagnetDriven2026, mondalSpinpolarizedAndreevMolecules2026,yangPerfectSpinJosephson2026}. However, most of these works do not explicitly invoke the fact that altermagnets are typically strongly spin-polarized materials, as compared to typical superconducting energy scales set by the superconducting gap. Here, we close that gap, emphasizing this fact and showing that the Josephson diode effect crucially depends on it.
 \begin{figure}[t!]
    \centering
    \includegraphics[width=1\linewidth]{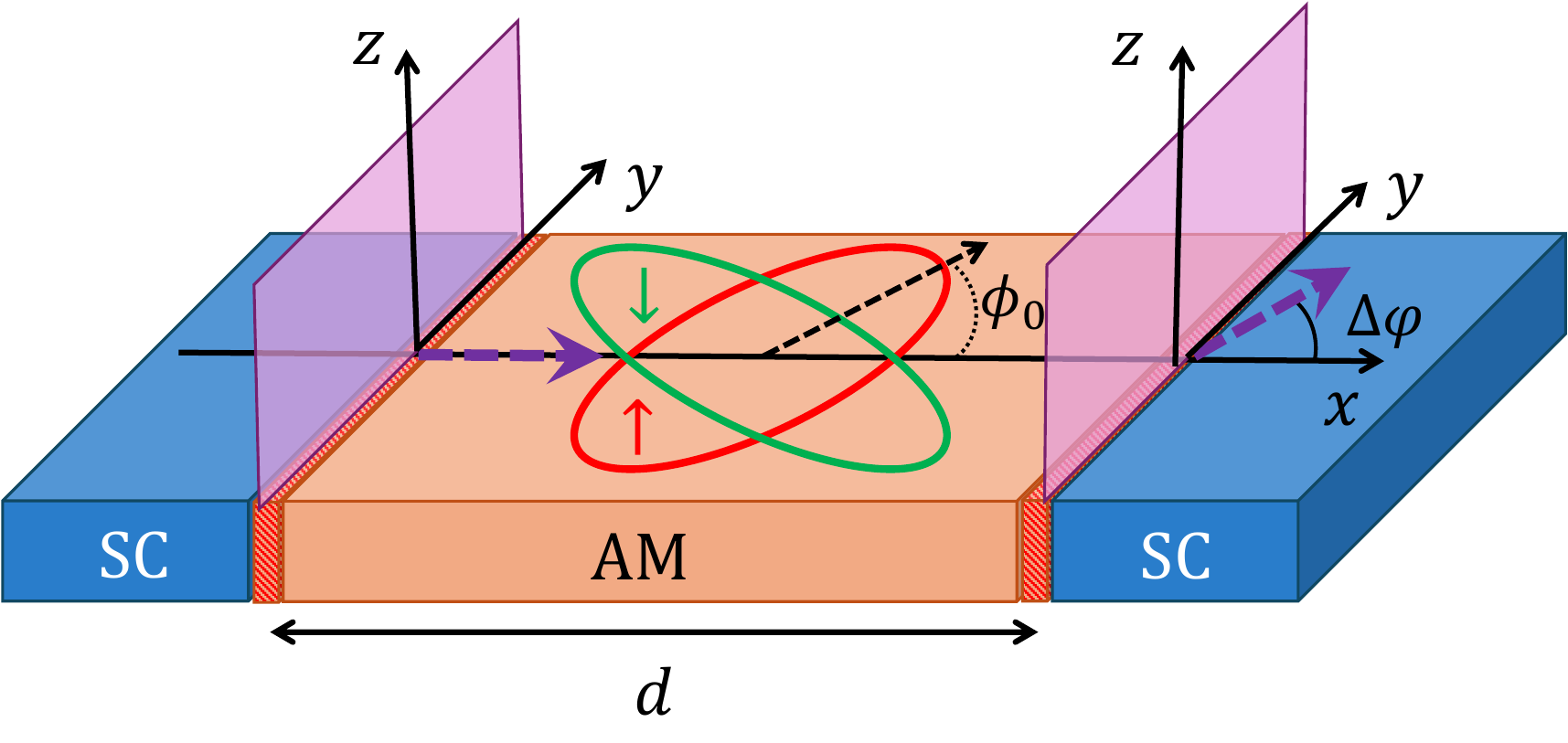}
    \caption{A scheme of the system under study consisting of an altermagnet (AM) coupled to two BCS superconductors (SC). In general, the SC/AM interfaces are spin-dependent with misaligned magnetic moments. For illustration, the altermagnet is shown in the $d_{xy}$ orientation ($\phi_0=\pi/4$); however, the results presented in the manuscript are obtained for the general case of an arbitrary orientation, $\phi_0\in[0,\pi/4]$ (see the text below). }
    \label{fig:system}
\end{figure}

In this paper, we discuss both the normal and anomalous Josephson effects, depending on whether the altermagnet is weakly or strongly spin-polarized. Based on the quantum Gor'kov and quasiclassical Eilenberger Green's function techniques, we account for the Josephson current-phase relation across the altermagnet coupled to two conventional superconductors, as shown in Fig.~\ref{fig:system}. The quasiclassical theory of superconductivity was already discussed in Refs.~\cite{giilQuasiclassicalTheorySuperconducting2024,kokkelerQuantumTransportTheory2025,herasInterplaySuperconductivityAltermagnetism2025}; however, only in weakly spin-polarized systems, which are characterized by one degenerated Fermi surface. Here, we develop the quasiclassical theory suited for strongly spin-polarized situations as well. Having obtained the Josephson current-phase relation,
we show that in the case of a weakly spin-polarized altermagnet, only the normal Josephson effect appears. On the contrary, in the opposite, strongly spin-polarized regime, the Josephson diode effect may appear. The main prerequisite for such an effect is the presence of a noncoplanar magnetization profile across the junction, which breaks inversion symmetry and is enabled by two thin ferromagnetic insulators located at the SC/AM interfaces. Such a magnetization profile leads to the emergence of the so-called quantum geometric phases, $\Delta\varphi$. This quantity is determined by the relative azimuthal angle between the exchange field vectors of the two ferromagnetic insulating layers (see Fig.~\ref{fig:system}), and it enters the Josephson CPR very similarly to the superconducting phase difference (for more details, see Refs.~\cite{greinSpinDependentCooperPair2009,schulz2025_prl,schulz2025_prb,nikolic_2025,nikolicNecessaryConditionsSpinresolved2025}). As a result, significant charge and spin Josephson diode effects emerge.

The article is organized as follows. In Sec.~\ref{Sec:weak} we briefly analyze the case of a weakly spin-polarized altermagnetic junction which displays a normal Josephson effect. In Sec.~\ref{Sec:strong} we present the results for a strongly spin-polarized altermagnetic junction which displays the anomalous charge and spin Josephson effects. In this section, we provide a detailed analysis of such effects, and the presented results represent the central message of this paper. In Sec.~\ref{sec:Discussion}, we discuss the results presented in the preceding section. Finally, in Sec.~\ref{sec:Method} we provide details on the quasiclassical Green's function applied to superconducting hybrid structures with unconventional magnets, both in a weakly and a strongly spin-polarized regime. 

\section{Weakly spin-polarized altermagnet: Normal Josephson effect}\label{Sec:weak}

We first apply our model to an SC/AM/SC heterostructure, considering a weakly spin-polarized altermagnet and fully transparent SC/AM interfaces. This regime assumes that the altermagnetic exchange field is much smaller than the Fermi energy, $\max_{\vec{p}_F}|h(\vec{p}_F)|=h_0\ll E_F$, which means that the quantum spin coherence between the two spin bands is maintained. Consequently, the system is described by the Eilenberger equation in the following form:
\begin{equation}\label{eqn:Eilenberger_2}
i\hbar\vec{v}_F\cdot\!\bm{\nabla}_\vec{R}\check{g}(\vec{R})+\qty[i\omega_n\check{\tau}_3+h(\vp_F)\check{\sigma}_3-\check{\Delta},\check{g}(\vec{R})]=0.
\end{equation}
 A detailed derivation of such an equation is given Sec.~\ref{Sec:theory_weak} [see in particular Eq.~\eqref{eqn:Eilenberger}]. Here, we just note that $\omega_n=(2n+1)\pi k_BT$ are Matsubara energies with $n\in\mathbb{Z}$ and temperature $T$, and $h(\vec{p}_F)$ is the altermagnetic exchange field. Note that in the superconductor $h\equiv 0$, whereas in the altermagnet $\check{\Delta}\equiv\check{0}$. Assuming the SC/AM interfaces are non-spin-active and considering a collinear altermagnet implies no magnetic inhomogeneity in the system. Consequently, equal-spin triplet correlations ($s_z=\pm 1$) do not exist in the system, and the quasiclassical Green's function has the following matrix form in particle-hole $\otimes$ spin space:
\begin{equation}
    \check{g}=\begin{pmatrix}
        g_{\uparrow\uparrow} & 0 & 0 & f_{\uparrow\downarrow} \\
        0 & g_{\downarrow\downarrow} & f_{\downarrow\uparrow} & 0\\
        0 & -\tilde{f}_{\uparrow\downarrow} & -\tilde{g}_{\uparrow\uparrow} & 0 \\
        -\tilde{f}_{\downarrow\uparrow} & 0 & 0 & -\tilde{g}_{\downarrow\downarrow}
    \end{pmatrix}.
\end{equation}
Utilizing this form of the quasiclassical propagator brings Eq.~\eqref{eqn:Eilenberger_2} into two decoupled spin-scalar Eilenberger equations:
\begin{equation}
    i\hbar\vec{v}_F\cdot\bm{\nabla}\hat{g}_\sigma(\vec{R})+\qty[\qty(i\omega_n+\zeta_\sigma h)\hat{\tau}_3-\zeta_\sigma\hat{\Delta},\hat{g}_\sigma(\vec{R})] = 0,
\end{equation}
where $\hat{g}_\sigma$ and $\hat{\Delta}$ have a $2\times 2$ matrix structure in particle-hole space only, and read, respectively,
\begin{equation}
    \hat{g}_\sigma=\begin{pmatrix}
        g_{\sigma\sigma} & f_{\sigma\bar{\sigma}} \\
        -\tilde{f}_{\bar{\sigma}\sigma} & -\tilde{g}_{\bar{\sigma}\bar{\sigma}}
    \end{pmatrix}~~\text{and}~~
    \hat{\Delta}=\begin{pmatrix}
        0 & \Delta e^{i\phi} \\
        -\Delta e^{-i\phi} & 0
    \end{pmatrix}.
\end{equation} 
Here, $\phi$ refers to the superconducting phase of the complex order parameter, $\bar{\sigma}$ is the spin sign opposite to $\sigma=\uparrow\downarrow$, and $\zeta_{\uparrow/\downarrow} = \pm 1$. The temperature dependence of the superconducting gap is taken from the well-known interpolation formula~\cite{muhlschlegelThermodynamischenFunktionenSupraleiters1959}
\begin{equation}
    \Delta(T)=\Delta_0\tanh\qty[1.74\sqrt{\frac{T_c}{T}-1}],
\end{equation}
where $\Delta_0$ is the gap value at $T=0$ and $T_c$ is the superconducting transition temperature. In what follows, we apply the above model to the SC/AM/SC junction depicted in Fig.~\ref{fig:system}.

Assuming a stepwise gap profile (no inverse proximity effect), the Eilenberger equation can be solved analytically for the system under consideration. For this purpose, we introduce two new Pauli matrices $\hat{\tau}_\phi=\mathrm{antidiag}(e^{i\phi},e^{-i\phi})$ and $\hat{\bar{\tau}}_\phi=i\hat{\tau}_\phi\hat{\tau}_3=\hat{\tau}_{\phi-\pi/2}$, which obey the Pauli algebra, $[\hat{\tau}_3,\hat{\tau}_\phi]=2i\hat{\bar{\tau}}_\phi$ and cyclic permutations, as well as $\hat{\tau}_3^2=\hat{\tau}_\phi^2=\hat{\bar{\tau}}_\phi^2=\hat{\tau}_0$ and $\{\hat{\tau}_3,\hat{\tau}_\phi \}=\{\hat{\tau}_3,\hat{\bar{\tau}}_\phi \} =\{\hat{\tau}_\phi,\hat{\bar{\tau}}_\phi\}=\hat{0}$. Using this basis, the general solution of Eq.~\eqref{eqn:Eilenberger_2} in the superconductor (where $h(\vec{p}_F)\equiv0)$ consists of a homogeneous solution, $\hat{g}^\mathrm{S}_{\sigma,h}$, and two evanescent modes, $\hat{g}^\mathrm{S}_{\sigma,\pm}$, which can be written as~\cite{nikolicLocalDensityStates2019}
\begin{align}
\hat{g}^{\mathrm{S}}_{\sigma,h}(\phi)&=\frac{1}{\Omega_n}(\omega_n\hat{\tau}_3-\zeta_\sigma\Delta\hat{\bar{\tau}}_\phi),\\
\hat{g}^{\mathrm{S}}_{\sigma,\pm}(x,\phi)&=\frac{1}{2\Omega_n}[\zeta_\sigma\Delta\hat{\tau}_3\mp i\Omega_n\hat{\tau}_\phi-\omega_n\hat{\bar{\tau}}_\phi]e^{\pm{2\Omega_nx}/{\hbar v_{Fx}}}\nonumber\\
&=\hat{g}_{\sigma,\pm}(\phi)e^{\pm{2\Omega_nx}/{\hbar v_{Fx}}}.
\end{align}
Here, we use the standard notation, $\Omega_n=\sqrt{\omega_n^2+\Delta^2}$ and $v_{Fx}=v_F\cos{\varphi_\vec{p}}$. Analogously, the general solution in the altermagnet (where $\Delta\equiv0$) reads
\begin{align}
    \hat{g}^\AM_{\sigma,h} &= \hat{\tau}_3,\\
    \hat{g}^\AM_{\sigma,\pm}(x)&=\hat{\tau}_\pm e^{\pm 2(\omega_n-i\zeta_\sigma h)x/\hbar v_{Fx}},
\end{align}
where $\hat{\tau}_\pm=\frac{1}{2}(\hat{\tau}\pm i\hat{\tau}_2)=\frac{1}{2}(\hat{\tau}_\phi\pm i\hat{\bar{\tau}}_\phi)e^{\pm i\phi}$.
To find a solution for the structure in Fig.~\ref{fig:system}, we make the following ansatz. We consider the trajectory starting in the left superconductor at $x=-\infty$ with $\phi=\chi_1$, going straight through the altermagnet of thickness $d$, ending up in the right superconductor at $x=\infty$ with $\phi=\chi_2$:
\begin{align}
    &\hat{g}_\sigma^\mathrm{S}(x<-d/2)=\hat{g}^\mathrm{S}_{\sigma,h}(\chi_1)\!+\!B_\sigma^+\hat{g}_{\sigma,+}(\chi_1)e^{{2\Omega_nx}/{\hbar v_{Fx}}}\!,\!\\
    &\hat{g}^\AM_\sigma(x)=A_\sigma\hat{g}^\AM_{\sigma,h} + A_\sigma^{-}\hat{g}^\AM_{\sigma,-}(x) + A_\sigma^+\hat{g}^\AM_{\sigma,+}(x),\\
    &\hat{g}_\sigma^\mathrm{S}(x>d/2)\!=\!\hat{g}^\mathrm{S}_{\sigma,h}(\chi_2)+B_\sigma^-\hat{g}_{\sigma,-}(\chi_2)e^{-{2\Omega_nx}/{\hbar v_{Fx}}}.
\end{align}
The unknown coefficients $A_\sigma,\,A_\sigma^\pm$, and $B_\sigma^\pm$ should be determined from the boundary conditions at $x=\pm d/2$. Note that the normalization of the Green's function $\hat{g}_\sigma^2=\hat{\tau}_0$ imposes the following relation: $A_\sigma^2+A_\sigma^+A_\sigma^-=1$. Assuming the ideal SC/AM interfaces, the quasiclassical Green's function is continuous throughout the system, yielding
\begin{align}
    \hat{g}^\mathrm{S}_\sigma(x=-d/2) &= \hat{g}^\AM_\sigma(x=-d/2),\\
    \hat{g}^\mathrm{S}_\sigma(x=d/2) &= \hat{g}^\AM_\sigma(x=d/2).
\end{align}
The above conditions result in a linear algebraic system of equations for the unknown coefficients, which can be solved analytically. As a result, the solution for the normal component of the quasiclassical Green's function, $g_{\sigma\sigma}$, in the altermagnet yields
\begin{equation}\label{eqn:g_weak}
    g^\AM_{\sigma\sigma}(i\omega_n,\vp_F) = \frac{\omega_n\cos\vartheta_\sigma(\vp_F)+i\Omega_n\sin\vartheta_\sigma(\vp_F)}{\Omega_n\cos\vartheta_\sigma(\vp_F)+i\omega_n\sin\vartheta_\sigma(\vp_F)}.
\end{equation}
\begin{figure}[t!]
    \centering
    \includegraphics[width=1\linewidth]{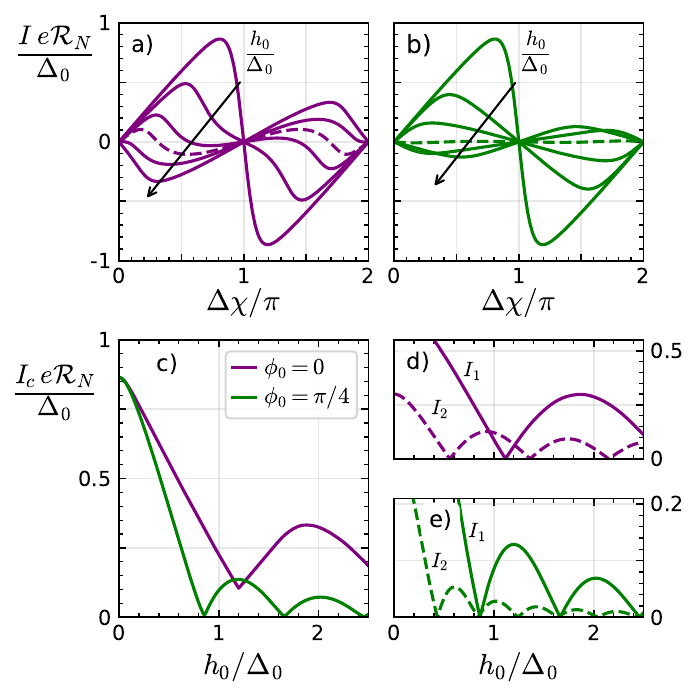}
    \caption{The Josephson CPR in an SC/AM/SC junction for altermagnet's thickness of $d=\xi$ at temperature $T=0.1\, T_c$, different values of altermagnetic exchange field $h_0/\Delta_0$, and (a) the $d_{x^2-y^2}$ orientation and (b) the $d_{xy}$ orientation. Panel (c): the corresponding critical currents $I_c$ as functions of the altermagnetic exchange field $h_0/\Delta_0$. Panels (d) and (e): the amplitudes of the first harmonic (solid lines) and the second harmonic (dashed lines) as functions of the altermagnetic exchange field $h_0/\Delta_0$ for the two orientations.}
    \label{fig:WeakReg_PerfectInt}
\end{figure}
We introduce the phase $\vartheta_\sigma$, which contains information about the altermagnetic field
\begin{equation}\label{eqn:theta}
    \vartheta_{\sigma}(\vp_F) = \frac{\Delta\chi}{2} - \frac{\zeta_\sigma h(\vp_F)d}{\hbar v_F\cos\varphi_\vp}-i\frac{\omega_n d}{\hbar v_F\cos\varphi_\vp},
\end{equation}
with $\Delta\chi=\chi_2-\chi_1$ being the superconducting phase difference across the junction and $\varphi_\vp$ the impact angle of the quasiclassical trajectory. Using this parametrization, the altermagnetic exchange field can be expressed in a compact form
\begin{equation}
    h(\vp_F,\phi_0) = h_0\cos(2\varphi_\vp-2\phi_0),
\end{equation}
where $\phi_0$ refers to the orientation of the altermagnet and $h_0$ is the strength of the altermagnetic exchange field. 
Introducing $\varphi_\vp$ and $\phi_0$ allows us to express the two characteristic orientations of the altermagnet compactly. Specifically, the $d_{xy}$ orientation is simply given by $h(\vp_F,\phi_0=\pi/4)=h_0\sin(2\varphi_\vp)$, whereas the $d_{x^2-y^2}$ orientation takes the form $h(\vp_F,\phi_0=0)=h_0\cos(2\varphi_\vp)$.  
The solution of Eq.~\eqref{eqn:g_weak} has a familiar form obtained for similar systems involving a homogeneous ferromagnet (obtained by setting $h(\vec{p}_F)=h_0=$ const., see Ref.~\cite{radovicCoexistenceStableMetastable2001}), a normal metal (obtained by setting $h_0=0$, see~\cite{nikolicLocalDensityStates2019}), or a spin-orbit material (by passing to helical basis, see Ref.~\cite{zhuangHelicalPhasesBogoliubov2025}). As one may notice from Eqs.~\eqref{eqn:g_weak} and~\eqref{eqn:theta}, the Green's function has two characteristic length scales: (i) the superconducting coherence length $\xi=\hbar v_F/\Delta_0$ and (ii) the exchange coherence length $\xi_h=\hbar v_F/h_0$. If the altermagnet is weakly spin polarized, $h_0\sim \Delta_0$, the two length scales are of similar order; therefore, from now on, we express the junction's length in units of the superconducting coherence length $\xi$. The supercurrent discussed below is expressed in units of $\Delta_0/(e\mathcal{R}_N)$, where the normal resistance $\mathcal{R}_N$ is defined as $\mathcal{R}_N=\pi^2\hbar/(e^2k_FW)$ with $W$ being the width in transverse ($y$-) direction. Note that our model assumes $W\gg \xi$.

Having obtained the quasiclassical propagator, the supercurrent density can be calculated as
\begin{equation}
    \bm{j}(\vec{R})\!=\!-ie\pi N_F k_BT\!\sum_{\sigma=\uparrow\downarrow}\!\sum_{n}\!\expval{\vec{v}_Fg_{\sigma\sigma}(\vec{p}_F,\vec{R},i\omega_n)}_{\vec{p}_F}\!,\!
\end{equation}
where $\expval{\bullet}_{\vec{p}_F}$ denotes the averaging over the Fermi surface [see Eq.~\eqref{eqn:Fermi_avg}] and we use the symmetry relation $\tilde{g}_{\bar\sigma \bar\sigma}=g_{{\sigma}\sigma}$. Note that due to a weak altermagnetic spin polarization, we assume identical and approximately spherical Fermi surfaces for both spin bands (see Supplemental Material). As we discuss in the subsequent section, this approximation cannot be done in the case of a strong spin polarization, which is crucial for the appearance of the anomalous Josephson effect. The resulting Josephson CPR is presented in Figs.~\ref{fig:WeakReg_PerfectInt}(a) and~\ref{fig:WeakReg_PerfectInt}(b), which show the $d_{x^2-y^2}~(\phi_0=0)$ and $d_{xy}~(\phi_0=\pi/4)$ orientations, respectively. Similarly to the ferromagnetic case, for certain values of the exchange field, the $d_{x^2-y^2}$-altermagnetic junction [see Fig.~\ref{fig:WeakReg_PerfectInt}(a)] undergoes the $0-\pi$ transition with a pronounced second harmonic due to the coexistence of the two phases~\cite{radovicCoexistenceStableMetastable2001}. However, the second harmonic obtained in this way is sensitive to the system parameters. To stabilize it, the presence of equal-spin triplets, and therefore a magnetic inhomogeneity in the system, is necessary. In contrast, the $d_{xy}$ orientation of the altermagnet shown in Fig.~\ref{fig:WeakReg_PerfectInt}(b) features a strongly suppressed second harmonic in the CPR close to the $0-\pi$ transition; a similarity to antiferromagnetic junctions~\cite{andersen0TransitionsJosephson2006}. To get a better insight, in Fig.~\ref{fig:WeakReg_PerfectInt}(c), we show the critical (maximum) currents for the $d_{x^2-y^2}$ orientation (violet line) and the $d_{xy}$ orientation (green line) as functions of the altermagnetic field strength. In the spirit of the above discussion, we note that the $d_{xy}$ orientation features almost zero current at the $0-\pi$ transition point, which clearly refers to a strongly suppressed second harmonic amplitude. On the contrary, the CPR in the case of the $d_{x^2-y^2}$ orientation is dominated by the second harmonic at the transition point. The corresponding first harmonic (solid lines) and second harmonic (dotted lines) amplitudes of the CPR are displayed in Figs.~\ref{fig:WeakReg_PerfectInt}(d) and ~\ref{fig:WeakReg_PerfectInt}(e), confirming the above conclusions. 

\section{Strongly spin-polarized altermagnet: Second harmonic generation and JDE}\label{Sec:strong}
In the preceding section, we discussed a junction involving a weakly spin-polarized altermagnet whose exchange field is comparable to the superconducting gap, i.e., much smaller than the Fermi energy. However, typically, altermagnets are strongly spin-polarized. As we show in the following, from the mesoscopic superconductivity perspective, this regime exhibits a variety of interesting phenomena, including the anomalous Josephson effect. Therefore, in this section, we put focus on a Josephson SC/AM/SC junction involving a strongly spin-polarized altermagnet, characterized by $\Delta\ll h_0\lesssim E_F$. Due to the strong splitting between the two spin bands, the quantum spin coherence is maintained only within each of them, but not between the two. In other words, such a system allows only for equal-spin triplet correlations ($\uparrow\uparrow$ and $\downarrow\downarrow$), and the altermagnet is described by two decoupled spin-scalar Eilenberger equations defined in particle-hole space only~[see Eq.~\eqref{eqn:strong_spin_Eilenberger}]
\begin{equation}
    i \hbar \mathbf{v}_{F,\sigma}\cdot \hat{g}_\sigma(\mathbf{R}) + \left[ i\omega_n \hat{\tau}_3 -\hat{\Sigma}_\sigma(\vec{R}), \hat{g}_\sigma(\mathbf{R}) \right] = 0.
    \label{eqn:strong_spin_Eilenberger_2}
\end{equation}
Here, 
\begin{equation}
    \hat{g}_\sigma=\begin{pmatrix}
        g_{\sigma\sigma} & f_{\sigma\sigma}\\
        -\tilde{f}_{\sigma\sigma} & - \tilde{g}_{\sigma\sigma}
    \end{pmatrix},
\end{equation}
and $\hat{\Sigma}_\sigma$ refers to the self-energy that contains the information about interactions in the system. Depending on the nature of the interaction, this self-energy can involve only one spin band, $\hat\Sigma_\sigma=\hat\Sigma_\sigma(\hat{g}_\sigma)$, or mixing between the two spin bands, $\hat\Sigma_\sigma=\hat\Sigma_\sigma(\hat{g}_\sigma,\hat{g}_{\bar{\sigma}})$. In what follows, we assume no interactions other than the altermagnetic exchange field in the altermagnet and the BCS pairing in the superconductors. A detailed derivation of Eq.~\eqref{eqn:strong_spin_Eilenberger_2} in given in Sec.~\ref{sec:Method}.

Utilizing the Riccati parametrization~\cite{Schopohl_Maki1995,schopohlTransformationEilenbergerEquations1998,eschrigDistributionFunctionsNonequilibrium2000}, and assuming no inverse proximity effect, the general solution of the above equation can be obtained analytically. Considering a particular mesoscopic setup, such as the junction sketched in Fig.~\ref{fig:system}, the full solution requires applying the appropriate boundary conditions. For this purpose, we follow Ref.~\cite{eschrigScatteringProblemNonequilibrium2009} and first formulate the boundary conditions solving the scattering problem at the SC/AM interface. Then we apply the iteration procedure, where the outgoing quasiclassical amplitudes at the one interface propagate throughout the altermagnet of length $d$, being viewed as the incoming quasiclassical amplitudes on the other interface. Each iteration is suppressed by a factor of $e^{-2|\omega_n|d/\hbar v_{F,\sigma}}$ compared to the previous one, ensuring a quick convergence after just a few iterations. The details on the construction of the S-matrix for an SC/AM interface and its implementation in the boundary condition problem are given in Supplemental Material. Once obtained, the quasiclassical propagator can be used for calculating the spin-resolved supercurrents as [see Eq.~\eqref{eqn:current_strongly}]
\begin{equation}\label{eqn:current_strongly_2}
    \bm{j}(\vec{R})\!=\!\frac{-ie\pi k_BT}{2}\!\sum_{\sigma,n}N_\sigma\!\Tr\expval{\vec{v}_\sigma\hat{\tau}_3\hat{g}_\sigma(i\omega_n, {\vec{p}_\sigma},\vec{R})}_{\sigma},
\end{equation}
where $\expval{\bullet}_\sigma$ refers to the averaging over the corresponding Fermi surface of $\sigma=\uparrow\!/\!\downarrow$ band. Note that in this case, the elliptical nature of the Fermi surfaces has to be taken into account. Moreover, the difference in momentum-resolved densities of states for the two spin bands is a prerequisite for the Josephson diode effect in strongly spin-polarized materials~\cite{nikolicNecessaryConditionsSpinresolved2025}.

As mentioned above, in strongly spin-polarized materials, transport is mediated by equal-spin triplets only. As it is known, the existence of these correlations is determined by the presence of magnetic inhomogeneity in the system. This regime can be realized either by an intrinsically inhomogeneous spin texture or by a non-local multilayer ferromagnetic geometry. Since the altermagnet considered here is collinear, we assume the latter scenario; namely, we allow for thin ferromagnetic insulating layers modeled as spin-dependent $\delta$ potentials positioned at the SC/AM interfaces (see Fig.~\ref{fig:system}). In what follows, we first consider the case of a coplanar exchange field profile across the junction, obtaining the normal Josephson effect. In particular, we show that for two spin-active SC/AM interfaces, the CPR hosts all harmonics (odd and even), indicating their long-range nature. On the other hand, keeping one interface (e.g., the left one) non-spin-active leads to a purely even-harmonic CPR, showing that only the so-called crossed-pair transmission processes occur. After this, we discuss a noncoplanar exchange field profile across the junction that leads to the anomalous Josephson effect and the Josephson diode effect. Since the interfacial ferromagnetic insulators are modeled by spin-dependent $\delta$ potentials, they are characterized by dimensionless BTK-like parameters~\cite{BTK1982}, $\mathcal{V}_i=2mV_i/(\hbar^2 k_F)$ and $\mathcal{J}_i=2mJ_i/(\hbar^2 k_F), (i=L,R)$. Most results are obtained with equal chemical potentials for the superconductors and the altermagnet, $\mu_\S=\mu_\AM=\mu$, the altermagnet thickness $d=\xi$, and $\mathcal{V}_i=0.2$ and $\mathcal{J}_i=0.1$, unless stated otherwise. 
\subsection{Coplanar magnetization profile: Normal Josephson effect and long-range second harmonic}\label{sec:strong_coplanar}
\begin{figure}[t!]
    \centering
    \includegraphics[width=1\linewidth]{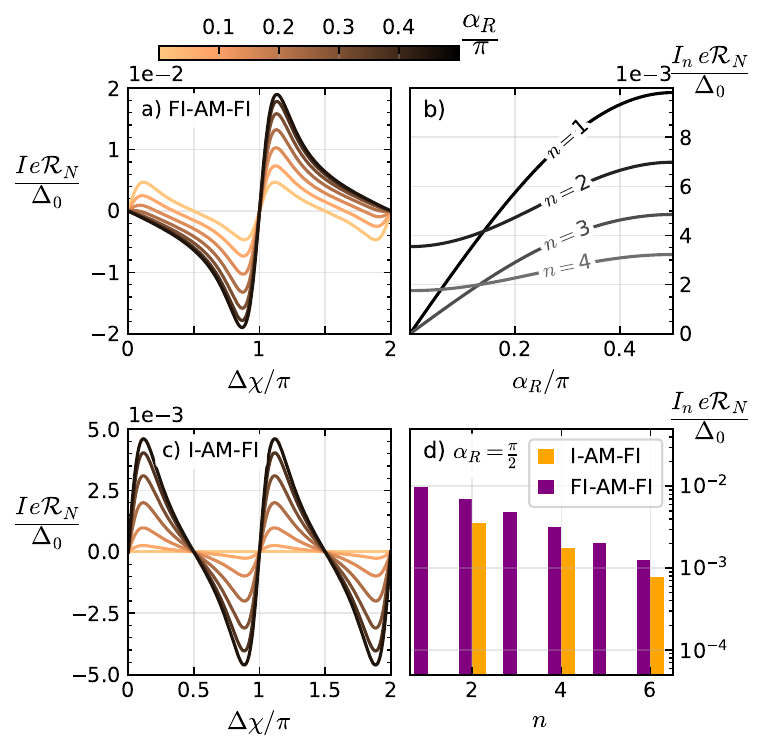}
    \caption{The Josephson CPR in different altermagnetic junctions. Panel (a) shows the Josephson CPR across an altermagnet coupled to superconductors via two spin-active interfaces ("trilayer") for different misalignment between the altermangetic exchange field $\vec{h}$ and the exchange field of the right SC/AM interface $\vec{J}_R$. Panel (b) shows first four harmonic amplitudes as a function of $\alpha_R$. Panel (c) shows the corresponding CPR in the system featuring one interface spin active and the other normal insulating ("bilayer"). Panel (d) shows the harmonic amplitudes in the case of a trilayer (purple) and a bilayer (orange).}
    \label{fig:coplanar}
\end{figure}

We first consider an altermagnet coupled to two superconductors via two spin-active interfaces, as shown in Fig.~\ref{fig:system}. Since we have three magnetic layers, we term this configuration a "trilayer". Below we will present the results for the case of one spin-active and another non-spin-active interface that we will term a "bilayer". For illustration, we examine the $d_{xy}$ orientation of the altermagnet since the existence of the effect presented in this subsection does not depend on the orientation. In the subsequent subsection, we will show that this is not the case in noncoplanar systems. Since we focus on the coplanar exchange field arrangement, we set the quantum geometric phase $\Delta\varphi$ to zero. Note that only a noncollinear (to be distinguished from noncoplanar) magnetization arrangement yields a nonvanishing supercurrent, since the Josephson transport in our model is mediated solely by the equal-spin correlations. Therefore, we set the exchange field of the left ferromagnetic insulating layer to be orthogonal to the exchange field vector of the altermagnet, $\alpha_L=\pi/2$ (considering the coordinate systems as in Fig.~\ref{fig:system}, it means $\vec{J}_L=J_L\vec{e}_x$). The misalignment between the two exchange field vectors ensures a nonvanishing current irrespective of the orientation of the exchange field of the right ferromagnetic insulator, parametrized by $\alpha_R$. However, the Josephson CPR itself is strongly affected by $\alpha_R$, as shown in Fig.~\ref{fig:coplanar}(a). The current is maximum for $\alpha_R=\pi/2$, which corresponds to parallel exchange fields of the two ferromagnetic insulators [both orthogonal to $\vec{h}(\vec{p}_F)$]. This maximization of the current is expected as both SC/AM interfaces allow for spin rotation, creating equal-spin triplet correlations. Decreasing $\alpha_R$ from $\pi/2$ to $0$ overall suppresses the critical current. However, it leads to the emergence of a pronounced second harmonic in the CPR. Moreover, the case of $\alpha_R=0$ leads to the purely even-harmonic CPR. To quantify this statement, we perform a harmonic expansion of the CPR
\begin{equation}\label{eqn:Fourier}
    I(\Delta\chi)=\sum_{n=1}^\infty I_n\sin(n\Delta\chi),
\end{equation}
and in Fig.~\ref{fig:coplanar}(b) we show the absolute values of the first four coefficients vs. the misalignment angle $\alpha_R$. We note that for parallel interfacial exchange field vectors, $\alpha_R=\pi/2$, the first harmonic, $n=1$, is dominant; however, with a significant contribution from the second harmonic, $n=2$. By decreasing $\alpha_R$, the fraction of even harmonics is getting more significant, so that $I_2$ becomes dominant for $\alpha_R\lesssim 0.1\pi$, while further decrease makes even $I_4$ stronger than $I_1$. Finally, for $\alpha_R=0$, odd harmonics are completely suppressed, and the CPR is fully even-harmonic. As already mentioned, in order to distinguish this configuration from the one presented below, we term it "trilayer", as we have three spin-active regions.

As shown above, if the exchange field of one of the interfacial ferromagnetic insulators aligns with that of the altermagnet, the Josephson CPR displays only even harmonics. This configuration motivates us to investigate the setup in which one SC/AM interface, e.g., the left one, is non-spin-active, $J_L=0$, keeping the other as before, $J_R\neq0$. This configuration we term a "bilayer", and the resulting Josephson CPR is displayed in Fig.~\ref{fig:coplanar}(c) for different values of $\alpha_R$ ranging from $\pi/2$ to $0$. Remarkably, the CPR contains only even harmonics. In addition, the critical current decreases monotonically with $\alpha_R$ until it completely diminishes for $\alpha_R=0$, due to the absence of equal-spin triplets in this regime. To quantify the effect, in Fig.~\ref{fig:coplanar}(d), we present the Fourier coefficients, $I_n$, obtained from Eq.~\eqref{eqn:Fourier} for the trilayer (violet bars) and the bilayer (orange bars). We note that the CPR of the former hosts all harmonics, whereas the CPR of the latter contains only even harmonics. In both cases, the lowest harmonics contribute the most to the CPR. We conclude that if the altermagnet is connected to the superconductors via two ferromagnetic insulating layers, all harmonics can be long-ranged. However, if the altermagnet is coupled to the superconductors via one ferromagnetic and one non-ferromagnetic interfacial insulating layer, only even harmonics are long-ranged. Similar effects have already been observed in ferromagnetic trilayers~\cite{braudeFullyDevelopedTriplet2007,eschrigTripletSupercurrentsClean2008,houzetFerromagneticJosephsonJunction2008,greinSpinDependentCooperPair2009,trifunovicLongrangeSpintripletProximity2010,schulz2025_prb,schulz2025_prl,nikolic_2025} and bilayers~\cite{pajovic_josephson_2006,trifunovic_josephson_2011,trifunovic_long-range_2011,melnikov_interference_2012,richard_superharmonic_2013,nikolicInterferencePhenomenaJosephson2022,schulzPurelyEvenHarmonic2026}.

\subsection{Noncoplanar spin arrangement: Josephson diode effect}\label{sec:strong_noncoplanar}
In this subsection, we analyze the effect of the noncoplanarity of the exchange field profile across the junction on the Josephson CPR. As before, we consider the system depicted in Fig.~\ref{fig:system}, but now allowing for nonzero $\Delta\varphi$. The noncoplanarity of the three exchange field vectors is determined by the condition $(\vec{J}_L\times \vec{J}_R)\cdot\vec{h}\neq 0$. Considering the altermagnetic field as a global spin quantization axis, we set $\vec{h}=h\vec{e}_z$ and the two additional exchange vectors we parametrize as $\vec{J}_i=J_i[\sin\alpha_i\cos\varphi_i,\sin\alpha_i\sin\varphi_i,\cos\alpha_i]^T$. The noncoplanarity condition reads $hJ_LJ_R\sin\alpha_L\sin\alpha_R\cos(\varphi_R-\varphi_L)\neq 0$. Since the existence of the equal-spin triplets, and consequently the Josephson current, already demands $\alpha_i\neq 0$, the noncoplanarity condition essentially depends on nonvanishing $\Delta\varphi=\varphi_R-\varphi_L$ that we term the \textit{quantum geometric phase difference}~\cite{greinSpinDependentCooperPair2009,schulz2025_prl,schulz2025_prb,nikolic_2025}. Since the existence of the Josephson diode effect does not depend on the specific values of $\alpha_i$, as long as they are nonzero, the results presented below are obtained for $\alpha_L=\alpha_R=\pi/2$. The other parameters take the values mentioned at the beginning of Sec.~\ref{Sec:strong} if not explicitly indicated otherwise in the captions of the figures and the text.

The model of a strongly spin-polarized system introduced above [see Eq.~\eqref{eqn:strong_spin_Eilenberger_2}] and elaborated in more detail in Sec.~\ref{Sec:theory_strong}  allows for an insight into spin-resolved quantities. In particular, we focus on the spin-resolved currents, $I_{\sigma}$ with $\sigma=\uparrow\downarrow$, calculated from Eq.~\eqref{eqn:current_strongly_2}. Having obtained $I_\sigma$, we easily express the charge and spin currents as $I_{\ch}=2e(I_\uparrow+ I_\downarrow)$ and $I_{\sp}=\hbar(I_\uparrow- I_\downarrow)$, respectively. In what follows, we put the main focus on the charge Josephson diode effect, accompanied by a brief discussion about the spin Josephson diode effect.  
\begin{figure}[t!]
    \centering
    \includegraphics[width=1\linewidth]{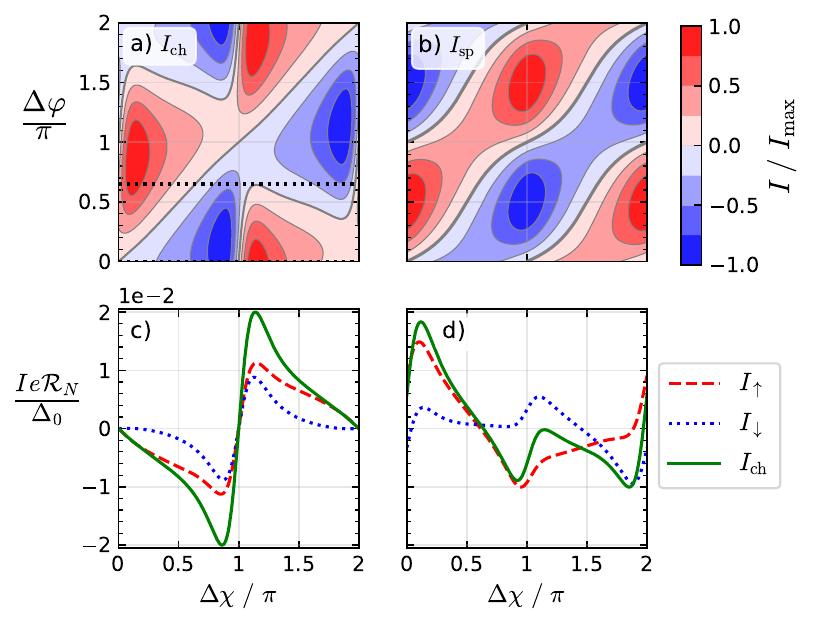}
    \caption{ Panel (a): The charge current $I_\ch$ as function of the superconducting phase difference $\Delta\chi$ and the quantum geometric phase $\Delta\varphi$. Panel (b): The corresponding spin current $I_\sp$. Horizontal cross-sections of the color scale plot from (a) taken for $\Delta\varphi=0$ and $\Delta\varphi=0.65\pi$ [see the dotted line in panel (a)] are shown in panels (c) and (d), respectively. The currents are calculated for the $\phi_0=0$ ($d_{x^2-y^2}$ orientation of the altermagnet), the temperature $T/T_c=0.1$, and the altermagnetic exchange field of $h_0/E_F=0.5$. Other parameters are mentioned at the beginning of the section.}
    \label{fig:CPR_00_orientation}
\end{figure}

Figure~\ref{fig:CPR_00_orientation}(a) shows the charge Josephson current $I_\ch$ as a function of the superconducting phase difference $\Delta\chi$ and the geometric phase difference $\Delta\varphi$ for the $d_{x^2-y^2}$ orientation of the altermagnet, $\phi_0=0$. Other parameters are indicated in the caption. Figure~\ref{fig:CPR_00_orientation}(b) shows the corresponding spin Josephson current $I_\sp$. Note that each quantity is normalized to the respective maximum value, namely, $I_{\ch}/I_{\ch,\mathrm{max}}$ and $I_{\sp}/I_{\sp,\mathrm{max}}$. Since we mainly focus on the charge CPR, in Figs.~\ref{fig:CPR_00_orientation}(c) and (d), we show the charge CPR represented by the solid green line, accompanied by the spin-resolved CRPs represented by the dashed and dotted lines. Figs.~\ref{fig:CPR_00_orientation}(c) displays the coplanar case, realized for $\Delta\varphi=0$, and discussed in Sec.~\ref{sec:strong_coplanar} [see, in particular, Fig.~\ref{fig:coplanar}(a)]. We note that the junction is in the $\pi$ state, exhibiting the normal Josephson effect for both spin-resolved currents and total charge current. This case corresponds to the horizontal cut of the plot in panel (a) taken at $\Delta\varphi=0$. For comparison, in Fig.~\ref{fig:coplanar}(d), we show the case of a noncoplanar spin arrangement, plotting the corresponding CPRs obtained for $\Delta\varphi=0.65\pi$ [see the dotted line in panel (a)]. Apparently, the spin-resolved currents as well as the total charge current display the difference in maximum values in the positive ($I_c^+$) and negative directions ($I_c^-$), also known as the Josephson diode effect. In the next subsection, we quantify this effect by calculating the so-called diode efficiency function and show that this case maximizes it. Here, we add two important remarks. First, we notice that the JDE is absent for $\Delta\varphi=n\pi$ with $n\in\mathbb{Z}$, since these cases correspond to coplanar magnetization profiles. However, in the case of $\Delta\varphi=\pi$, the junction undergoes the $\pi-0$ transition. Second, in Fig.~\ref{fig:CPR_00_orientation} we present the results for the $d_{x^2-y^2}$ orientation of the altermagnet, which maximizes the effect. The reason for this lies in the fact that this orientation yields the most significant half-metallic contribution necessary for the effect~\cite{greinSpinDependentCooperPair2009,schulz2025_prl,schulz2025_prb,nikolic_2025}. In the Supplemental Materials, we present the results for two additional orientations. As we show, in the case of $\phi_0=\pi/8$ the effect is weaker, while in the case of $\phi_0=\pi/4$ (orientation $d_{xy}$) the effect completely diminishes due to the absence of the half-metallic contribution. 

\subsection{Charge and spin diode efficiency}
In the following, we investigate the Josephson diode effect more quantitatively. The figure of merit in this effect is the so-called diode efficiency, typically defined as
\begin{equation}\label{eqn:eta}
    \eta_\alpha = \dfrac{|I_\alpha^+| - |I_\alpha^-|}{|I_\alpha^+|+|I_\alpha^-|} \ ,
\end{equation}
where $\alpha=\ch/\sp$ refers to the charge/spin. In addition, $I_\alpha^\pm \equiv I_\alpha(\Delta\chi_\mathrm{ch}^\pm)$, where $\Delta\chi^\pm$ are the superconducting phase differences at which the positive or negative critical \textit{charge current} are reached for fixed $\Delta\varphi$, i.e., $\Delta\chi_\mathrm{ch}^+=\mathrm{argmax}_{\Delta\chi}(I_\mathrm{ch})$ and $\Delta\chi_\mathrm{ch}^-=\mathrm{argmin}_{\Delta\chi}(I_\mathrm{ch})$. In what follows, we will first focus on the charge JDE with a significant efficiency of more than $30\%$ in the particular setup. Then we will comment on the spin JDE, showing that the junction can exhibit a perfect spin diode effect of $\eta_\sp=100\%$.

\begin{figure}[t!]
    \centering
    \includegraphics[width=1\linewidth]{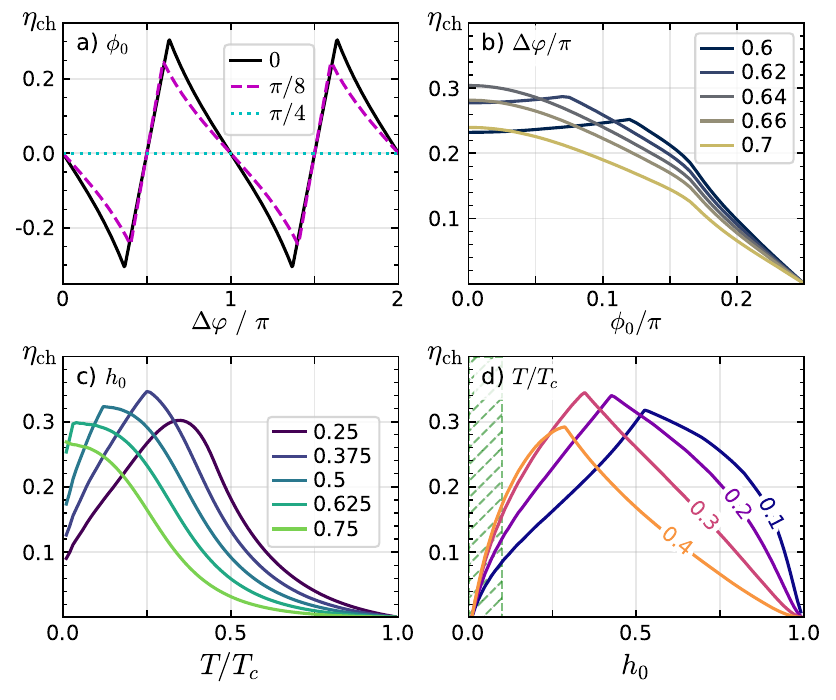}
    \caption{Charge diode efficiency $\eta_\mathrm{ch}$ as a function of various system's parameters. Panel (a) shows the dependence of $\eta_ch$ on the quantum-geometric angle $\Delta \varphi$ for three different orientations of the altermagnet given by $\phi_0$. Panel (b) shows the dependence of the orientation of the altermagnet for different geometric angles. Panel (c) shows the role of temperature $T$ for different altermagnetic exchange strengths $h_0$. Panel (d) shows the dependence on the altermagnetic exchange fields for different temperatures.}
    \label{fig:charge_diode_efficiencies}
\end{figure}

 In Fig.~\ref{fig:charge_diode_efficiencies}, we present the charge diode $\eta_\mathrm{ch}$ against different parameters of the system. Panel (a) shows $\eta_\ch$ as a function of the quantum geometric phase $\Delta\varphi$ for three different altermagnetic orientations. In all cases, the JDE vanishes for $\Delta \varphi = k \frac{\pi}{2},\ k \in \mathbb{Z}$ since the Josephson CPR in this case possesses a phase-inversion center~\cite{nikolicNecessaryConditionsSpinresolved2025}; this effect is in accordance with previous works on ferromagnetic junctions~\cite{schulz2025_prl,schulz2025_prb,nikolic_2025}. Overall, the effect is the strongest for $\phi_0=0$ ($d_{x^2-y^2}$ orientation; solid black line), with a maximum value of $\eta_\ch\approx 33.7\%$ achieved for $\Delta\varphi_0\approx 0.65\pi+k\pi$. The corresponding CPRs were shown in Fig.~\ref{fig:CPR_00_orientation}(d). As already mentioned, the $d_{x^2-y^2}$ orientation maximizes the half-metallic contribution to the Josephson current, making the JDE maximum as well. Shifting towards the $d_{xy}$ orientation shifts the maximum of $\eta_\ch(\Delta\varphi)$ and, overall, suppresses the effect; the case of $\phi_0=\pi/8$ still displays a considerable JDE, while the effect diminishes for $\phi_0=\pi/4$. The effect of the orientation of the altermagnet is explicitly visible in Fig.~\ref{fig:charge_diode_efficiencies}(b), where we show the charge diode efficiency $\eta_\ch$ versus $\phi_0$ for different quantum geometric phases $\Delta\varphi$. We note that for a fixed value of $\Delta\varphi$ the effect can be even enhanced if the altermagnet is rotated away from the $d_{x^2-y^2}$ position, $\phi_0>0$. However, the maximum efficiency is a monotonically decreasing function of the orientation, and the effect always diminishes for the $d_{xy}$ case. Temperature effects are presented in Fig.~\ref{fig:charge_diode_efficiencies}(c), where we plot the charge diode efficiency $\eta_\ch$ as a function of temperature $T$ for different values of the altermagnetic exchange field $h_0$. For strong spin polarization of the altermagnet ($h_0=0.75 E_F$; see the green line), $\eta_\ch$ is a monotonically decreasing function of $T$. However, for intermediate altermagnetic field strengths, the effect can be even enhanced by temperature. Decreasing $h_0$ shifts the maximum diode efficiency towards higher temperatures. For instance, the exchange field of $h_0=0.25 E_F$ shifts the maximum $\eta_\ch$ towards rather high temperatures $T>0.3T_c$. Finally, in Fig.~\ref{fig:charge_diode_efficiencies}(d), we present the effect of the altermagnetic field strength $h_0$, plotting the $\eta_\ch(h_0)$ function for a few values of $T$. Similarly to the effect of $T$, the charge diode efficiency is not a monotonic function of $h_0$. The maximum efficiency is typically achieved for some intermediate values of $h_0$ around $0.4E_F$. Increasing temperature tends to shift the maximum $\eta_\ch$ towards lower values of $h_0$. Note that for $h_0=E_F$ the effect vanishes since we have no altermagnetic ordering anymore. Also, the green hatched area in Fig.~\ref{fig:charge_diode_efficiencies}(d) refers to the regime where our model is not quantitatively valid anymore, since the strong spin polarization is required. This domain is better described by the quasiclassical theory in the weakly spin-polarized regime formulated in Sections~\ref{Sec:weak} and~\ref{Sec:theory_weak}. However, in sufficiently long junctions, $d\gg\xi$, where the mixed-spin correlations are substantially suppressed as compared to the equal-spin ones, the above conclusions remain qualitatively valid.

\begin{figure}[t!]
    \centering
    \includegraphics[width=1\linewidth]{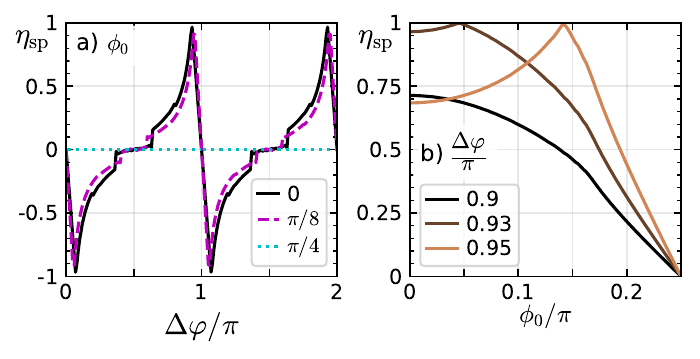}
    \caption{The spin diode efficiency $\eta_\sp$ as a function of: (a) the quantum geometric phase $\Delta\varphi$ for three orientations of the altermagnet and (b) the altermagnetic orientation $\phi_0$ for three values of the geometric phase. System's parameters are as in Fig.~\ref{fig:charge_diode_efficiencies}.}
    \label{fig:spin_diode_efficiency}
\end{figure}

For comparison, in Fig.~\ref{fig:spin_diode_efficiency} we present the results for the spin diode efficiency obtained for the same system discussed above. As already mentioned in Eq.~\eqref{eqn:eta}, the spin diode efficiency is obtained as
\begin{equation}
     \eta_\sp = \dfrac{|I_\sp(\Delta\chi^+)| - |I_\sp(\Delta\chi^-)|}{|I_\sp(\Delta\chi^+)| + |I_\sp(\Delta\chi^-)|} \ ,
\end{equation}
where $\Delta\chi^\pm$ are the superconducting phase differences at which the positive or negative \textit{critical charge current} is reached for a particular set of system parameters. Figure~\ref{fig:spin_diode_efficiency}(a) shows the spin diode efficiency $\eta_\sp$ as a function of the quantum geometric phase $\Delta\varphi$ for the same parameters as in Fig.~\ref{fig:charge_diode_efficiencies}(a). Remarkably, close to $\Delta\varphi=n\pi, n\in\mathbb{Z}$ the cases of $\phi_0=0$ ($d_{x^2-y^2}$ orientation) and $\phi_0=\pi/8$ exhibit nearly the perfect spin diode efficiency of $\eta_\sp=100\%$. Note that this small discrepancy is due to a particular set of parameters. In other words, the perfect spin diode effect is not prohibited by fundamental reasons, as one can see in Fig.~\ref{fig:spin_diode_efficiency}(b), which shows the $\eta_\sp(\phi_0)$ function for different values of $\Delta\varphi$. We note that the cases of $\Delta\varphi=0.93\pi$ and $\Delta\varphi=0.95\pi$ display the perfect spin diode effect for particular orientations. Finally, similarly to the charge diode effect discussed above, the $d_{xy}$ orientation ($\phi_0=\pi/4$)  of the altermagnet displays no spin diode effect. 
\subsection{Harmonic analysis of the spin-resolved currents}
As shown above, in the case of a strongly spin-polarized altermagnet, the two spin bands decouple, maintaining the pairing coherence within each band but not between the two. As a result, the Josephson transport is mediated by two distinct spin channels, allowing for insight into spin-resolved supercurrents, $I_\sigma$ with $\sigma=\uparrow\downarrow$. Under time reversal, these supercurrents transform as
\begin{equation}
    I_\sigma(\Delta\chi,\Delta\varphi) = -I_\sigma(-\Delta \chi, -\Delta\varphi),
\end{equation}
following from the behavior of the Hamiltonian under time reversal. Namely, considering that the altermagnetic exchange field $\vec{h}(\vec{p}_F)$ originates from a mean-field treatment of spin Heisenberg interactions, it reverses its direction under time reversal. Consequently, the spin quantization axis that is set by the unit vector in the direction of $\vec{h}$ also reverses under time reversal. With this convention, we keep the assignment between majority/minority spin and spin-up/spin-down intact. More details on this matter can be found in Ref.~\cite{nikolicNecessaryConditionsSpinresolved2025}. Since the presence of higher harmonics in the CPR is of vital importance for the Josephson diode effect, a detailed harmonic analysis is of interest. Numerical study of the spin-resolved CPRs presented in the previous subsections suggests the following harmonic expansion in $\Delta\chi$ and $\Delta\varphi$:
\begingroup
\allowdisplaybreaks
\begin{align}
    I_{\uparrow} &= \frac{1}{2}\sum_{\mu,\nu=-\infty}^\infty  (-1)^{\mu+\nu} \mu I_{\mu,\nu} \sin\psi_{\mu,\nu}, \label{eqn:I_uu} \\ 
    I_{\downarrow} &= \frac{1}{2}\sum_{\mu,\nu=-\infty}^\infty  (-1)^{\mu+\nu} \nu I_{\mu,\nu} \sin\psi_{\mu,\nu}, \label{eqn:I_dd}
\end{align}
\endgroup
where $I_{-\mu,-\nu}=I_{\mu,\nu}$, and 
\begin{equation}\label{eqn:effJos}
\psi_{\mu,\nu} = (\mu+\nu)\Delta\chi - (\mu-\nu)\Delta\varphi
\end{equation}
is the effective Josephson phase. Such an expansion was proposed for the first time in Ref.~\cite{eschrigSpinpolarizedSupercurrentsSpintronics2015}, and further developed in Refs.~\cite{schulz2025_prl,schulz2025_prb,nikolic_2025}, having an appealing and intuitive physical interpretation. Namely, Eqs.~\eqref{eqn:I_uu} and~\eqref{eqn:I_dd} represent the Josephson CPRs mediated by a coherent transfer of $\mu$ $\uparrow\uparrow$ and $\nu$ $\downarrow\downarrow$ equal-spin pairs. Note that $\mu$ and $\nu$ are algebraic, namely, their signs refer to the direction of propagation. Given the fact that the ground state of the junction is a $\pi$-state [see Fig.~\ref{fig:CPR_00_orientation}(c)], the total phase acquired by a single $\uparrow\uparrow\!/\!\downarrow\downarrow$ pair is  $\Delta\chi\mp\Delta\varphi+\pi$. Therefore, the process involving $\mu$ $\uparrow\uparrow$ pairs and $\nu$ $\downarrow\downarrow$ pairs is accompanied by the total phase of $\mu(\Delta\chi-\Delta\varphi+\pi)+\nu(\Delta\chi+\Delta\varphi+\pi)=(\mu+\nu)\Delta\chi-(\mu-\nu)\Delta\varphi+(\mu+\nu)\pi$, leading to Eqs.~\eqref{eqn:I_uu}-\eqref{eqn:effJos}.

\begin{figure} [t!]
    \centering
    \includegraphics[width=1\linewidth]{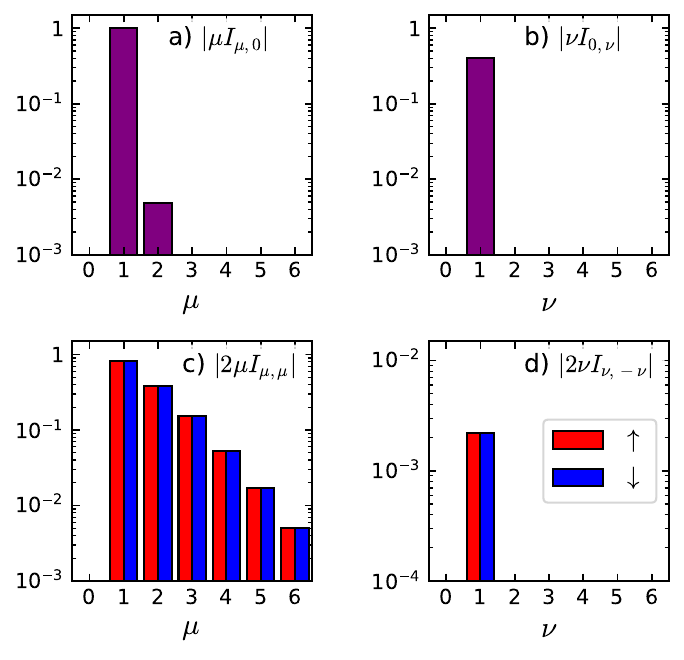}
    \caption{Harmonic decomposition of the spin-resolved Josephson CPRs. Panels (a) and (b) show $I_{\mu,0}$ and $I_{0,\nu}$ amplitudes, respectively. The former corresponds to the coherent transfer of $\mu$ $\uparrow\uparrow$ pairs and zero $\downarrow\downarrow$ pairs, whereas the latter describes the coherent transfer of $\nu$ $\downarrow\downarrow$ pairs and zero $\uparrow\uparrow$ pairs. Panels (c) and (d) show the crossed pair contributions $I_{\mu\mu}$ and $I_{\mu,-\mu}$, respectively. The red/blue color corresponds to spin-$\uparrow$/spin-$\downarrow$.}
    \label{fig:harmonics_00}
\end{figure}
To illustrate Eqs.~\eqref{eqn:I_uu} and \eqref{eqn:I_dd}, we consider the Josephson CPRs obtained for the $d_{x^2-y^2}$ orientation of the altermagnet and described in Sec.~\ref{sec:strong_noncoplanar}. {Based on the above discussion, one may expect that the processes involving fewer Cooper pairs are contributing more to the current. Therefore, we start the discussion by considering processes with only one spin band being involved. Fig.~\ref{fig:harmonics_00}(a) shows the amplitudes for the process mediated only by the $\mu$ $\uparrow\uparrow$ pairs, namely, $\mu I_{\mu,0}$ in Eq.~\eqref{eqn:I_uu}. We note that the dominant process involves the transfer of just one Cooper pair, while the higher processes are greatly suppressed. Considering transport mediated by the spin-$\downarrow$ band, in Fig.~\ref{fig:harmonics_00}(b) we present the $\nu I_{0,\nu}$ coefficients. Similarly to the spin-$\uparrow$ band, here the dominant process is the one involving just one Cooper pair, i.e., $I_{0,1}$; higher processes are even more strongly suppressed than in Fig.~\ref{fig:harmonics_00}(a).} 

{Considering processes that involve both spin bands, the especially important contributions to the JDE are those coming from the so-called crossed-pair transmission processes; namely, the ones in which an equal number of $\uparrow\uparrow$ and $\downarrow\downarrow$ pairs is transferred, as pointed out in Ref.~\cite{schulz2025_prb, schulz2025_prl,nikolic_2025,nikolicNecessaryConditionsSpinresolved2025}. Figure~\ref{fig:harmonics_00}(c) shows the $\mu I_{\mu,\mu}$ coefficients corresponding to the coherent transfer of $\mu~\uparrow\uparrow$ pairs and $\mu~\downarrow\downarrow$ pairs in the same direction. The red (blue) bars correspond to the $\uparrow\uparrow~(\downarrow\downarrow)$ case. One may note that these coefficients are the same for both spin bands, numerically confirming the expansion of Eqs.~\eqref{eqn:I_uu} and~\eqref{eqn:I_dd}, and that $\mu I_{\mu\mu}$ decreases approximately exponentially with $\mu$. Analogously, Fig.~\ref{fig:harmonics_00}(d) shows the ${\nu I_{\nu,-\nu}}$ coefficients corresponding to the coherent propagation of ${\nu}~\uparrow\uparrow$ pairs and ${\nu}~\downarrow\downarrow$ pairs in the opposite directions. For symmetric junctions, such processes overall are significantly suppressed as compared to $\mu I_{\mu,\mu}$. In addition, they do not contribute to the charge supercurrent, but only to the spin supercurrent. Note that all coefficients in Fig.~\ref{fig:harmonics_00} are normalized to {$I_{1,0}$} and their absolute values are shown. Here, we have not commented on the practical numerical procedure of calculating the above coefficient. Instead, we refer the reader to Refs.~\cite{schulz2025_prb,nikolic_2025} that contain a detailed discussion on this matter.}
\section{Discussion}\label{sec:Discussion}
In the preceding sections, we have presented a systematic theoretical model for the normal Josephson effect and the Josephson diode effect in mesoscopic proximity structures consisting of a d-wave altermagnet placed between two BCS superconductors. The model has been formulated in the framework of the quantum Gor'kov and the quasiclassical Eilenberger Green's functions techniques. Our theory is adapted to two distinct regimes: (i) a weakly spin-polarized altermagnet whose exchange field is considerably smaller than the Fermi energy and comparable to the superconducting gap, and (ii) a strongly spin-polarized altermagnet whose exchange field is comparable to the Fermi energy. 

Considering a weak spin polarization of the altermagnet, we have obtained a reciprocal Josephson current-phase relation; however, strongly affected by the orientation of the altermagnet. Namely, the $d_{x^2-y^2}$ altermagnetic orientation features a strong second harmonic in the CPR close to the $0-\pi$ transition; the effect is similar to that in ferromagnetic junctions. On the contrary, the $d_{xy}$ orientation features almost no second harmonic close to the transition point, which is similar to the behavior of antiferromagnetic junctions. In both cases, the junction exhibits characteristic oscillations of the critical current with the thickness or the strength of the altermagnetic exchange field due to $0-\pi$ transitions. 

Considering a strongly spin-polarized regime, we have distinguished two sub-regimes. First, we have considered a coplanar exchange field profile allowed by the presence of two thin ferromagnetic insulating layers positioned at the SC/AM interfaces. In this case, the normal Josephson effect appears due to the preservation of the inversion symmetry. In the case of two spin-active SC/FM interfaces, all harmonics in the Josephson CPR are long-ranged. However, if we set one SC/AM interface to be non-magnetic and keep the other magnetic, we have shown that the Josephson CPR features only even harmonics. As discussed in Ref.~\cite{schulzPurelyEvenHarmonic2026} for the case of ferromagnetic junctions, this effect is attributed to higher-order Andreev processes occurring at the non-magnetic SC/AM interface.  

Finally, we have considered the noncoplanar exchange field profile across the junctions, showing the appearance of the Josephson diode effect in the CPR. Our particular set of parameters has led to a charge diode efficiency of $\eta_\ch\approx 32\%$ and a perfect spin diode of $\eta_\sp=100\%$. However, the value of $\eta_\ch\approx 32\%$ is not bounded by some fundamental physical reasons, and a more careful exploration of the parameter space may yield even higher diode efficiencies. However, we stress that the effect is maximum for the $d_{x^2-y^2}$ orientation of the altermagnet due to the significant half-metallic contribution caused by the mismatch in the Fermi surface geometries of the altermagnet and the superconductor. On the contrary, due to the same reason, the $d_{xy}$ altermagnetic orientation yields no diode effect at all. To quantify our findings, we have performed the harmonic analysis of the spin-resolved Josephson CPRs, which has allowed us to describe the effects presented in our work in terms of the coherent transfer of multiple equal-spin triplet pairs across the altermagnet. 
\section{Method}\label{sec:Method}
We model a structure of clean two-dimensional films as displayed in Fig.~\ref{fig:system}, where an altermagnet (AM; orange) is placed between two s-wave BCS superconductors (SC; blue). In general, we allow for thin ferromagnetic insulating layers at SC/FM interfaces modeled by spin-dependent $\delta$ functions. All metallic parts (superconductors and altermagnet) are described within the quasiclassical theory of superconductivity. However, on the contrary, the SC/AM interfaces are, in general, beyond the quasiclassical approximation, but they can be incorporated in the theory via the appropriate boundary conditions formulated in Refs.~\cite{eschrigScatteringProblemNonequilibrium2009,greinSpinDependentCooperPair2009}. In what follows, we first discuss a single-particle Hamiltonian describing an altermagnet and showing how the construction of the S-matrix is carried out. Note that the S-matrix is necessary for the formulation of the boundary condition problem in our model. Then we provide details on the many-body Green's function description, both in the quantum (Gor'kov) and quasiclassical (Eilenberger) regimes used for calculating the Josephson current. More details on the diagonalization of the altermagnetic Hamiltonian and implementation of the S-matrix in the boundary condition problem are given in Supplemental Material.
\subsection{Hamiltonian of the system}\label{Method:Hamiltionian}
Assuming that in the atomic vicinity of an SC/AM interface, the superconductor is in the normal state, the proximity contact can be described by the following single-electron Hamiltonian:
\begin{equation}\label{eqn:Hamiltonian}
\hat{H}(\vec{k}) = \frac{\hbar^2\vec{k}^2}{2m} - \mu + \hat{V}-h(\vec{k})\hat{\sigma}_3,
\end{equation}
where $h(\vec{k})$ refers to the altermagnetic exchange field given by~\cite{smejkalConventionalFerromagnetismAntiferromagnetism2022,smejkalEmergingResearchLandscape2022,Sun2023}
\begin{equation}\label{eqn:AM_field}
    h(\vec{k})\!=\!{h_0}\theta(x)\qty[\cos(2\phi_0)(\hat{k}_x^2\!-\!\hat{k}_y^2)\!+\!2\sin(2\phi_0)\hat{k}_x\hat{k}_y].
\end{equation}
In this equation, $h_0$ is the strength of the altermagnetic exchange field, here in units of energy, and $\hat{k}_i=k_i/k_F$ is a $k_i$ projection (where $i=x,y,z$) of the $k$-vector normalized to the Fermi wave vector $k_F=\sqrt{2m\mu/\hbar^2}$ in the normal state, $h_0=0$. The $\phi_0$ angle determines the orientation of the altermagnet, representing the crossover between the $d_{x^2-y^2}$ orientation obtained for $\phi_0=0$ and the $d_{xy}$ orientation obtained for $\phi_0=\pi/4$. For illustration, Fig.~\ref{fig:system} shows the $d_{xy}$ orientation of the altermagnet. Note that $\theta(x)$ refers to the Heaviside step function introduced for the reason that the altermagnet occupies the right half-space of the contact $(x>0)$. For definiteness, we assume equal chemical potential of the two regions $\mu$, and $\hat{V}$ describes the interfacial potentials modeled as
\begin{equation}
    \hat{V}(x) = \sum_{i=L,R}(V_{i}\hat\sigma_0 -J_{i}\mathbf{n}_{i}\cdot \hat{\boldsymbol{\sigma}})\delta(x-x_i),
\end{equation}
with the local spin quantization axes being parametrized by $\mathbf{n}_i=(\sin\alpha_i\cos\varphi_i,\sin\alpha_i\sin\varphi_i,\cos\alpha_i)^T$. As already mentioned, a nonvanishing $J_i$ is a prerequisite for the existence of equal-spin triplet correlations, $\uparrow\uparrow$ and $\downarrow\downarrow$, responsible for the Josephson diode effect in strongly spin-polarized altermagnets (see Sec.~\ref{Sec:strong}). However, such an assumption is not demanded in weakly spin-polarized altermagnets that feature the normal Josephson effect mainly mediated by the mixed-spin correlations, $\uparrow\downarrow\pm\downarrow\uparrow$ (see Sections~\ref{Sec:theory_weak} and~\ref{Sec:weak}).

Written in the form of Eq.~\eqref{eqn:AM_field}, the altermagnetic term is not Hermitian due to the presence of the $\theta$ function. To fix this and make it Hermitian, we need to symmetrize it. One possible way to do so is to rewrite it in the following form:
\begin{equation}
\begin{split}
    h(\vec{k}) = {h_0}\big[&\sin(2\phi_0) \{ \hat{k}_x, \theta(x)\}\hat{k}_y+\\
    &+\cos(2\phi_0) ( \hat{k}_x \theta(x) \hat{k}_x - \hat{k}_y \theta(x) \hat{k}_y)\big],
\end{split}
\end{equation}
where $\{\bullet\,,\bullet\}$ refers to the anticommutator. This form of the Hamiltonian is utilized for calculating the S-matrix that connects the waves from the two sides of the SC/AM interface, and that enters the boundary conditions for the quasiclassical propagators at the interface. Since this task is a straightforward and lengthy algebraic problem, we do not show the details here, but we refer the reader to the Supplemental Materials. Here, we stress that the S-matrix has a general $4\times4$ matrix structure in channel$\otimes$spin space: 
\begin{equation}\label{eqn:S_matrix}
    S = \begin{pmatrix}
        \hat{R}_1 & |T_{12}\rangle & |T_{13}\rangle \\
        \langle T_{21}| & r_{22} & r_{23} \\
        \langle T_{31}| & r_{32} & r_{33} 
    \end{pmatrix} \ ,
\end{equation}
with the following notation. The index 1 denotes the {doubly} spin-degenerate conduction channel on the superconducting side, whereas the indices $2$ and $3$ refer, respectively, to the $\uparrow$ and $\downarrow$ spin bands in the altermagnet. The hat and bracket notation indicate spin-space matrix or vector structure. For solving the boundary condition problem with two SC/AM interfaces (see Fig.~\ref{fig:system}), we have employed the 
iteration procedure. The outgoing amplitudes at one interface are propagated across the AM region and considered as the incoming amplitudes at the other interface until the convergence with the desired numerical accuracy is reached. Once obtained, the quasiclassical propagator is inserted in Eq.~\eqref{eqn:current_strongly} to calculate the Josephson current-phase relation. 
\subsection{Many body description: Gor'kov equations}\label{Method:Gorkov}

The cornerstone of the Green's function approach to mesoscopic superconductivity is the so-called Gor’kov Green’s function formalism, formulated in particle-hole space. Dealing with an additional spin degree of freedom (e.g., due to spin-dependent interactions) invokes the larger particle-hole $\otimes$ spin space. In the case of an altermagnet characterized by the anisotropic exchange field $h(\vec{p})$, the Gor’kov Green’s function obeys the following left Dyson-Gor'kov equation~\cite{gorkovEnergySpectrumSuperconductors1958,sereneQuasiclassicalApproachSuperfluid1983,eschrigSpinpolarizedSupercurrentsSpintronics2015}:
\begin{equation}\label{eqn:left_gorkov}
   [i\omega_n\check{\tau}_3\!-\!\xi(\hat{\vec{p}})\!+\!h(\hat{\vec{p}})\check{\sigma}_3\!-\!{\check{\underline{\Sigma}}\check{\tau}_3}]\otimes \check{\tau}_3\check{G}(\vec{r},\vec{r}')=\delta(\vec{r}-\vec{r}').
\end{equation}
Here, $\omega_n=(2n+1)\pi k_BT$ are fermionic Matsubara energies with $n\in\mathbb{Z}$ and temperature $T$, and $\xi(\hat{\vec{p}})=E(\hat{\vec{p}})-\mu$ is the dispersion relation with $\hat{\vec{p}}=-i\hbar\hat{\bm{\nabla}}$ being the canonical momentum and $\mu$ the chemical potential. Note that altermagnetic ordering has to be modeled by an even parity exchange field, $h(\hat{\vec{p}})=h(-\hat{\vec{p}})$. Throughout the paper, when we mention the strength of the altermagnetic field, we refer to $h_0=\max_\vp |h(\vp)|$. The self-energy function {$\check{\underline{\Sigma}}(\vec{p},\vec{r})$} contains various interacting terms, including the BCS pairing potential. More details on the specific form of $\check{\Sigma}$ are given depending on the context. Finally, $\hat{\tau}_i$ and $\hat{\sigma}_i$ are the Pauli matrices in Gor'kov-Nambu and spin space, respectively, and check symbol ($\check{\bullet}$) denotes a matrix in combined space, i.e., $\check{\tau}_i=\hat{\tau}_i\otimes\hat{\sigma}_0$ and $\check{\sigma}_i=\hat{\tau}_0\otimes\hat{\sigma}_i$. The Gor'kov Green's function is built from the basis {$\check{\Psi}=(\psi_\uparrow,\psi_\downarrow,\psi^\dagger_\uparrow,\psi^\dagger_\downarrow)$  having a $4\times 4$ matrix structure in combined space
\begin{equation}\label{eqn:GF}
    \check{G} = \left(
    \begin{array}{cccc}
         \hat{G} & \hat{F}\\
         \hat{\tilde{F}}& \hat{\tilde{G}}
    \end{array}{}
    \right)
    \quad\text{with}\quad
    \hat{F} = \left(
    \begin{array}{cccc}
         F_{\uparrow\uparrow} & F_{\uparrow\downarrow}\\
         F_{\downarrow\uparrow}& F_{\downarrow\downarrow}
    \end{array}{}
    \right).
\end{equation}}

Since the full Gor'kov approach is typically not practical for treating inhomogeneous mesoscopic systems, our objective here is to develop a quasiclassical theory~\cite{larkinQuasiclassicalMethodTheory1969,belzigQuasiclassicalGreensFunction1999}. For this purpose, we follow the standard procedure (see, e.g., Ref.~\cite{sereneQuasiclassicalApproachSuperfluid1983}), and first decompose the spatial dependence of the Gor'kov Green's function into the center-of-mass, $\vec{R}=(\vec{r}+\vec{r}')/2$, and the relative coordinate, $\vec{x}=\vec{r}-\vec{r}'$, contribution. Then we apply the gradient expansion to Eq.~\eqref{eqn:left_gorkov} up to the linear order in gradients in $\vec{R}$. The momentum-dependent part of Eq.~\eqref{eqn:left_gorkov} transforms as
\begin{align}
    &[\xi(\hat{\vec{p}})-h(\hat{\vec{p}})\check{\sigma}_3]\otimes \check{\tau}_3\check{G}\qty(\vec{R}+\frac{\vec{x}}{2},\vec{R}-\frac{\vec{x}}{2}) \approx \\
    &\approx\qty[\xi(\hvp)-h(\hvp)\check{\sigma}_3-\frac{i}{2}\qty(\frac{\partial\xi}{\partial\vp}-\frac{\partial h}{\partial\vp}\check{\sigma}_3)\bm{\nabla}_\vec{R}]\nonumber\check{\tau}_3\check{G}(\vp,\vec{R}).
\end{align}
Analogously, the right Dyson{-}Gor'kov equation reads
\begin{equation}\label{eqn:right_gorkov}
    \check{\tau}_3\check{G}(\vec{r},\vec{r}')\otimes[i\omega_n\check{\tau}_3-\xi(\hat{\vec{p}})+h(\hat{\vec{p}})\check{\sigma}_3-{\check{\underline{\Sigma}}\check{\tau}_3}] =\delta(\vec{r}-\vec{r}'),
\end{equation}
with the corresponding gradient expansion
\begin{align}
    &\check{\tau}_3\check{G}\qty(\vec{R}+\frac{\vec{x}}{2},\vec{R}-\frac{\vec{x}}{2})\otimes[\xi(\hat{\vec{p}})-h(\hat{\vec{p}})\check{\sigma}_3] \approx \\
    &\approx\check{\tau}_3\check{G}(\vp,\vec{R}) \qty[\xi(\hvp)-h(\hvp)\check{\sigma}_3+\frac{i}{2}\qty(\frac{\partial\xi}{\partial\vp}-\frac{\partial h}{\partial\vp}\check{\sigma}_3)\bm{\nabla}_\vec{R}]\nonumber.
\end{align}
Consequently, the left-right-subtracted Dyson{-}Gor'kov equation, obtained by subtracting Eqs.~\eqref{eqn:left_gorkov} and \eqref{eqn:right_gorkov}, reads:
{\begin{align}
    &\frac{i}{2}\bigg\{\qty(\hbar\vec{v}-\frac{\partial h}{\partial\vp}\check{\sigma}_3)\bm{\nabla}_\vec{R},\check{\tau}_3\check{G}(\vec{p},\vec{R})\bigg\}+\\ &\qquad\qquad+\qty[i\omega_n\check{\tau}_3\!+\!h(\hvp)\check{\sigma}_3-\check{\underline{\Sigma}}(\vec{R})\check{\tau}_3,\check{\tau}_3\check{G}(\vec{p},\vec{R})]=0,\nonumber
\end{align}}
where $\vec{v}(\vp)=\frac{\partial \xi}{\partial\vp}$ is the group velocity and $\{\bullet,\bullet\}$ denotes the anticommutator. 

The Green's function can be parametrized in a standard manner~\cite{sereneQuasiclassicalApproachSuperfluid1983}
\begin{equation}
    \check{\tau}_3\check{G}=\left(
    \begin{array}{cc}
        G_0\hat{\sigma}_0+\bm{G}\cdot\bm{\hat{\sigma}} &  (F_0+\bm{F}\cdot\bm{\hat{\sigma}})i\hat{\sigma}_2\\
     (\tilde{F}_0+\bm{\tilde{F}}\cdot\bm{\hat{\sigma}}^\ast)i\hat{\sigma}_2    & \tilde{G}_0\hat{\sigma}_0+{\bm{\tilde{G}}\cdot\bm{\hat{\sigma}^\ast}} 
    \end{array}{} \right),
\end{equation}
identifying the mixed- and the equal-spin contributions to it, respectively,
\begin{equation}\label{eqn:G_MS}
\check{\tau}_3\check{G}_{\rm MS}\!=\!\left(
    \begin{array}{cc}
        G_1\hat{\sigma}_1+G_2\hat{\sigma}_2 & (F_0\hat{\sigma}_0+F_3\hat{\sigma}_3)i\hat{\sigma}_2 \\
       (\tilde{F}_0\hat{\sigma}_0+\tilde{F}_3\hat{\sigma}_3) i\hat{\sigma}_2  & \tilde{G}_1\hat{\sigma}_1-\tilde{G}_2\hat{\sigma}_2
    \end{array}\right),
\end{equation}
\begin{equation}\label{eqn:G_ES}
\check{\tau}_3\check{G}_{\rm ES}\!=\!\left(
    \begin{array}{cc}
        G_0\hat{\sigma}_0+G_3\hat{\sigma}_3 & (F_1\hat{\sigma}_1+F_2\hat{\sigma}_2)i\hat{\sigma}_2 \\
       (\tilde{F}_1\hat{\sigma}_1-\tilde{F}_2\hat{\sigma}_2) i\hat{\sigma}_2  & \tilde{G}_0\hat{\sigma}_0+\tilde{G}_3\hat{\sigma}_3
    \end{array}\right).
\end{equation}
Keeping in mind that $[\check{\sigma}_3,\check{\tau}_3\check{G}_\mathrm{ES}]=\{\check{\sigma}_3,\check{\tau}_3\check{G}_\mathrm{MS}\}=0$ and       $[\check{\sigma}_3,\check{\tau}_3\check{G}_\mathrm{MS}]=2\check{\sigma}_3{\check{\tau}_3}\check{G}_\mathrm{MS}$ and $\{\check{\sigma}_3,\check{\tau}_3\check{G}_\mathrm{ES}\}=2\check{\sigma}_3{\check{\tau}_3}\check{G}_\mathrm{ES}$, we obtain
\begin{align}
    \label{eqn:MS}
&i\hbar\vec{v}\cdot\bm{\nabla}\check{\tau}_3\check{G}_\mathrm{MS}\!+\!\qty[i\omega_n\check{\tau}_3\!+\!h\check{\sigma}_3\!-{\check{\Sigma}_\mathrm{MS}\check{\tau}_3},\check{\tau}_3\check{G}_\mathrm{MS}]=0,\!\\
   \label{eqn:ES}
    &i\qty(\!\hbar\vec{v}\!-\!\frac{\partial h}{\partial\vp}\check{\sigma}_3\!)\!\cdot\!\bm{\nabla}\bm\check{\tau}_3\check{G}_\mathrm{ES}\!+\!\qty[i\omega_n\check{\tau}_3\!-\!{\check{\Sigma}_\mathrm{ES}\check{\tau}_3},\check{\tau}_3\check{G}_\mathrm{ES}]=0.
\end{align}
These equations lay the groundwork for the quasiclassical theory that we formulate in the two limiting cases: (i) weakly spin-polarized altermagnets, $h_0\sim\Delta\ll E_F$ (see Sec.~\ref{Sec:weak}), and (ii) strongly spin-polarized altermagnets, $\Delta\ll h_0\lesssim E_F$ (see Sec.~\ref{Sec:strong}). Here, $\Delta$ refers to the superconducting order parameter and $E_F$ to the Fermi energy, while $h_0$ is the strength of the altermagnetic field introduced above. Since we aim at describing superconductor (SC) - altermagnet (AM), superconductivity in the former is included via the BCS self-energy, 
{\begin{equation}\label{eqn:gap}
    \check{\underline{\Sigma}}\check{\tau}_3=\check{\Delta}=\begin{pmatrix}
        0 & \hat\Delta\\
        \tilde{\Delta} & 0
    \end{pmatrix},
\end{equation}}
where $\hat{\Delta}=\Delta i\hat{\sigma}_2$ and $\tilde{\Delta}=\hat{\Delta}^\ast$, which correspond to spin-singlet pairing that we focus on. By far, we have not fully applied the quasiclassical approximation, and in what follows, we consider the two limiting cases separately. {Note that the quasiclassical approximation applied in the following subsection assumes the multiplication of the above self-energy by $\check{\tau}_3$ from the right. For ease of notation, we will not introduce a new symbol for the quasiclassical self-energies, but from now on assume $\check{\underline{\Sigma}}(\vec{p},\vec{r})\check{\tau}_3\to\check{\Sigma}(\vec{p}_F,\vec{R})$.}
\subsection{Quasiclassical theory of S/AM interfaces with weakly spin-polarized AMs}\label{Sec:theory_weak}
Considering the case of a weakly spin-polarized altermagnet, $h_0\ll E_F$, Eqs.~\eqref{eqn:MS} and \eqref{eqn:ES} can be joined adopting a compact form 
\begin{equation}
i\hbar\vec{v}\cdot\bm{\nabla}_\vec{R}\check{\tau}_3\check{G}+\qty[i\omega_n\check{\tau}_3+h(\vp)\check{\sigma}_3-\check{\Sigma},\check{\tau}_3\check{G}]=0.
\end{equation}
Note that we have neglected the $\partial h/\partial\vp$ term in Eq.~\eqref{eqn:ES} due to $h_0\ll E_F\implies$ $|\partial h/\partial\vp|\ll|\vec{v}|$. Now we can apply the quasiclassical approximation~\cite{larkinQuasiclassicalMethodTheory1969,belzigQuasiclassicalGreensFunction1999}. Setting momentum to the Fermi momentum, $\vec{p}\to\vec{p}_F$ (and, correspondingly, $\vec{v}\to\vec{v}_F$), multiplying the whole expression by $i/\pi$ and integrating it over $d\xi_\vp$ close to the Fermi surface, we arrive at the\textit{ Eilenberger equation}~\cite{eilenbergerTransformationGorkovsEquation1968}
\begin{equation}\label{eqn:Eilenberger}
i\hbar\vec{v}_F\cdot\bm{\nabla}_\vec{R}\check{g}(\vec{R})+\qty[i\omega_n\check{\tau}_3+h(\vp_F)\check{\sigma}_3\!-\!\check{\Sigma},\check{g}(\vec{R})]=0.
\end{equation}
The notation is the same as before, and $\check{g}$ represents the quasiclassical Gor'kov Green's function defined as
\begin{equation}
    \check{g}(\vec{p}_F,\vec{R},i\omega_n)=\frac{i}{\pi}\oint d\xi_\vp\check{\tau}_3\check{G}(\xi_\vp;\vec{p}_F,\vec{R},i\omega_n).
\end{equation}
The quasiclassical propagator is normalized, and we choose the convention $\check{g}^2=\check{\tau}_0$, with $\check{g}$ having the following matrix structure in combined particle-hole $\otimes$ spin space:
\begin{equation}
  \check{g} = \begin{pmatrix}
      \hat{g} & \hat{f}\\
      -\hat{\tilde{f}} & -\hat{\tilde{g}}
  \end{pmatrix}\quad\text{with} \quad
  \hat{f}=\begin{pmatrix}
      f_{\uparrow\uparrow} & f_{\uparrow\downarrow}\\
      f_{\downarrow\uparrow} & f_{\downarrow\downarrow}
  \end{pmatrix}.
\end{equation}
{The BCS self energy (superconducting gap) in the quasiclassical approximation has the matrix form of Eq.~\eqref{eqn:gap}.} Note that in the expressions above tilde ($\tilde{\bullet}$) symbol denotes the particle-hole conjugation operation
\begin{equation}\label{eqn:tilde_operation}
    \tilde{Q}(\vp_F,\vec{R},i\omega_n) = Q^\ast(-\vp_F,\vec{R},i\omega_n).
\end{equation}
Having obtained $\check{g}$, we can express observables in a compact manner. For instance, the supercurrent density reads~\cite{sereneQuasiclassicalApproachSuperfluid1983,kopnin}
\begin{equation}\label{eqn:j}
    \bm{j}(\vec{R})\!=\!\frac{-ie\pi N_F k_BT}{2}\!\sum_{n}\!\Tr_4\langle\vec{v}_F\check{\tau}_3\check{g}(\vec{p}_F,\vec
    R,i\omega_n\!)\rangle_{\vec{p}_F},\!
\end{equation}
{Here,} the trace is performed over the combined space and $\langle\bullet\rangle_{\vec{p}_F}$ refers to the averaging over the Fermi surface,
\begin{subequations}\label{eqn:Fermi_avg}
\begin{eqnarray}
    &&\expval{\bullet} = \frac{1}{N_F}\int_\mathrm{FS_\sigma}\frac{d^{{d-1}}p_F}{(2\pi\hbar)^{{d}}|\vec{v}_F(\vec{p}_F)|}(\bullet),\\
    &&N_F = \int_\mathrm{FS_\sigma}\frac{d^{{d-1}}p_F}{(2\pi\hbar)^{{d}}|\vec{v}_F(\vec{p}_F)|},
\end{eqnarray}
\end{subequations}
{with $d$ referring to dimensionality. In this work, we assume a 2D altermagnet, meaning $d=2$.} Equation~\eqref{eqn:j} has been used for calculating the Josephson current across a weakly spin-polarized altermagnet coupled to two BCS superconductors presented in Sec.~\ref{Sec:weak}. Note that due to a weak spin-polarization in this regime, we assume a spin-degenerate, nearly spherical Fermi surface; more details can be found in the Supplemental Material.  
\subsection{Quasiclassical theory of S/AM interfaces with strongly spin-polarized AMs}\label{Sec:theory_strong}
Similarly to ferromagnetic contacts, in the case of an S/AM contact, it can be shown that mixed-spin correlations (spin singlets and spin triplets with $s_z=0$) penetrate the AM over considerably shorter length scales than equal-spin correlations (spin triplets with $s_z=\pm 1$). Namely, the characteristic decaying length of mixed-spin correlations is set by the ferromagnetic (exchange) coherence length $\sim\hbar v_F/h_0$, while the corresponding length scale for equal-spin triplets is given by the thermal coherence length $\sim \hbar v_F/(k_BT)$. We immediately notice that for $h_0\lesssim E_F$ the former decay on very small length scales comparable to the Fermi wavelength, $\lambda_F$, which is much shorter than the thermal coherence length, $\lambda_F\ll \hbar v_F/(k_BT)$. As a result, in a strongly spin-polarized altermagnet, $\Delta\ll h_0\lesssim E_F$, the mixed-spin correlations are nonvanishing only in an atomically thin region close to the S/AM interface. Consequently, they can be neglected compared to the equal-spin correlations on mesoscopic distances. From a technical point of view, this means that we drop Eq.~\eqref{eqn:MS} and keep only Eq.~\eqref{eqn:ES}. In other words, due to the strong splitting between the two spin bands, the coherence is maintained only within each band but not between them. In addition, in this case, the $|\partial h/\partial\vec{p}|$ term is not negligible as compared to $\vec{v}$, meaning that the two spin bands effectively have different band-specific Fermi velocities and momentum-resolved densities of states. This allows us to apply the quasiclassical approximation to each Fermi surface separately, arriving at two spin-scalar Eilenberger equations
\begin{equation}
    i \hbar \mathbf{v}_{F,\sigma}\cdot \hat{g}_\sigma(\mathbf{R}) + \left[ i\omega_n \hat{\tau}_3 -\hat{\Sigma}_\sigma(\vec{R}), \hat{g}_\sigma(\mathbf{R}) \right] = 0,
    \label{eqn:strong_spin_Eilenberger}
\end{equation}
where $\sigma=\uparrow\downarrow$ denotes the spin projection. Note that $\hat{g}_\sigma$ is now a $2\times2$ matrix in particle-hole space only, defined as
\begin{align}
    \hat{g}_\sigma(\mathbf{p}_{F,\sigma},\mathbf{R};i\omega_n) &=  \dfrac{i}{\pi}\int d\xi_{\mathbf{p},\sigma} \hat{\tau}_3 \hat{G}_\sigma(\mathbf{p}_{F,\sigma},\mathbf{R};i\omega_n)=\nonumber\\
    &=\begin{pmatrix}
        g_{\sigma\sigma} & f_{\sigma\sigma}\\
        -\tilde{f}_{\sigma\sigma} & -\tilde{g}_{\sigma\sigma}
    \end{pmatrix}.
\end{align}
As before, the quasiclassical propagators are normalized, and we take $\hat{g}_\sigma^2=\hat{\tau}_0$. In addition, the following symmetry is fulfilled: $\tilde{g}_{\sigma\sigma}=g_{\sigma\sigma}$. The normalization condition allows for choosing an appropriate parametrization of $\hat{g}_\sigma$. For the description of the system in the strongly spin-polarized regime, we employ the so-called Riccati parametrization, in which~\cite{Schopohl_Maki1995,schopohlTransformationEilenbergerEquations1998,eschrigDistributionFunctionsNonequilibrium2000}
\begin{equation}\label{eqn:gamma_parametrization}
    \hat{g}_\sigma=\begin{pmatrix}
        g_\sigma & f_\sigma\\
        -\tilde{f}_\sigma & -g_\sigma
    \end{pmatrix} = \frac{1}{1\!-\!\gamma_\sigma\tilde{\gamma}_\sigma}\begin{pmatrix}
        1+\gamma_\sigma\tilde{\gamma}_\sigma & 2\gamma_\sigma\\
        -2\tilde{\gamma}_\sigma & {-}\gamma_\sigma\tilde{\gamma}_\sigma-1
    \end{pmatrix}.
\end{equation}
Assuming no impurity or some other scattering processes, $\hat\Sigma_\sigma=0$, the Eilenberger equations in the altermagnet are greatly simplified, having the form
\begin{equation}\label{eqn:Ricatti_F}
  i\hbar\vec{v}_\sigma\cdot\bm{\nabla}\gamma_\sigma + 2i\omega_n\gamma_\sigma=0,  
\end{equation}
with a solution
\begin{equation}\label{eqn:gamma_AM}
    \gamma_\sigma(i\omega_n,x) = A_\sigma\ \mathrm{exp}\left[ -\dfrac{2\omega_n}{\hbar \mathbf{v}_{F,\sigma}\cdot \mathbf{e}_x}x \right].
\end{equation}
Considering a superconductor, the situation is different due to the spin coherence, so that the $\gamma$'s are not simply spin-scalars but matrices in spin space, $\hat\gamma_s$. The Eilenberger equation in this case reads
\begin{equation}\label{eqn:Riccati_S}
    (i\hbar\vec{v}_F\cdot\bm{\nabla}+2i\omega_n)\hat{\gamma}_s + \hat{\Delta} - \hat{\gamma}_s\tilde{\Delta}\hat{\gamma}_s=0.
\end{equation}
The results in Sec.~\ref{Sec:strong} are obtained for the case of a homogeneous superconductor, $\bm{\nabla}\hat\gamma_s=0$, for which the solution of the above equation is simply given by
\begin{equation}\label{eqn:gamma_S}
       \hat{\gamma}_s(i\omega_n) = \frac{-\Delta\hat{\sigma}_2}{\omega_n+\sqrt{\omega_n^2+|\Delta|^2}}.
\end{equation}
The corresponding tilde-quantities are obtained by applying the particle-hole conjugation [see Eq.~\eqref{eqn:tilde_operation}], arriving at
\begin{align}\label{eqn:tilde_gamma_AM}
    &\tilde\gamma_\sigma(i\omega_n,x) = B_\sigma\ \mathrm{exp}\left[ \dfrac{2\omega_n}{\hbar\mathbf{v}_{F,\sigma}\cdot \mathbf{e}_x}x \right],\\
    &\tilde{\hat{\gamma}}_s(i\omega_n) = \frac{\Delta^\ast\hat{\sigma}_2}{\omega_n+\sqrt{\omega_n^2+|\Delta|^2}}.
\end{align}
Note that $\gamma_\sigma$ is an electron-like propagator corresponding to the propagation along the trajectory parametrized by $\vec{p}_\sigma$, whereas $\tilde\gamma_\sigma$ corresponds to hole propagation in the opposite direction. The unknown coefficients $A_\sigma$ and $B_\sigma$ in Eqs.~\eqref{eqn:gamma_AM} and~\eqref{eqn:tilde_gamma_AM} are determined from the boundary conditions imposed at the S/AM interface. The details of the practical implementation of the boundary conditions are given in the Supplemental Materials. 

Having obtained the coherence amplitudes $\gamma_\sigma$ and $\tilde\gamma_\sigma$, we can calculate the Josephson current across the altermagnet as
\begin{equation}\label{eqn:current_strongly}
    \bm{j}(\vec{R})\!=\!\frac{-ie\pi k_BT}{2}\!\sum_{\sigma,n}N_\sigma\!\Tr\expval{\vec{v}_\sigma\hat{\tau}_3\hat{g}_\sigma(i\omega_n, {\vec{p}_\sigma},\vec{R})}_{\sigma},
\end{equation}
where $\hat{g}_\sigma$ is given by Eq.~\eqref{eqn:gamma_parametrization} and $\expval{\bullet}_\sigma$ refers to averaging over the corresponding Fermi surface, $\mathrm{FS}_\sigma$, with a full analogy to Eqs.~\eqref{eqn:Fermi_avg}. Equation~\eqref{eqn:current_strongly} has been used for studying the Josephson diode effect in Sec.~\ref{Sec:strong}.
\vspace{0.5cm}

\section*{Data availability}
The data sets generated in this study are available from the corresponding author upon reasonable request.
\section*{Code availability}
The codes used for obtaining the numerical results in this study are available from the corresponding author upon reasonable request.
\section*{Acknowledgments}
We thank W. Belzig and C. Wiedemann for illuminating discussions. This work was partly supported by Deutsche Forschungsgemeinschaft (DFG, German Research Foundation) under Project No. 530670387. The computations were enabled by resources provided by the University Computer Centre of the University of Greifswald.
\section*{Author contributions}
D.N. and M.E. conceptualized the research topic and supervised. J.F.N. performed numerical simulations and generated figures. J.F.N. and D.N. wrote the first version of the manuscript. All authors wrote the final version of the manuscript. 
\section*{Competing interests}
The authors declare no competing interests. 

\section*{Additional information}
The supplementary information for this study is available.

\bibliography{references}
\end{document}


\begin{center}\Large{Supporting Information for}\end{center}
\title{Spin-polarized supercurrents and Josephson diode effect in altermagnets}
\author{Janus F. Niebuhr}
\affiliation{Institut f\"ur Physik, Universit\"at Greifswald, Felix-Hausdorff-Strasse 6, 17489 Greifswald, Germany}
\author{Matthias Eschrig}
\affiliation{Institut f\"ur Physik, Universit\"at Greifswald, Felix-Hausdorff-Strasse 6, 17489 Greifswald, Germany}
\author{Danilo Nikoli\'c}
\affiliation{Institut f\"ur Physik, Universit\"at Greifswald, Felix-Hausdorff-Strasse 6, 17489 Greifswald, Germany}
\date{\today}

\begin{abstract} 
\end{abstract}
\maketitle
\newpage
\section{S-matrix for an SC/AM interface}\label{appendix:S_matrix}
As mentioned in the main text [see Eq.~(28)], we consider a system consisting of a superconductor SC $(x<0)$ and an altermagnet AM $(x>0)$ coupled via a thin spin-active insulator at the interface $(x=0)$. Assuming that the superconductor in an atomically thin region close to the interface is in the normal state the system is described by the following single-electron Hamiltonian~\cite{smejkalConventionalFerromagnetismAntiferromagnetism2022,smejkalEmergingResearchLandscape2022}
\begin{equation}
    \hat{H}(\mathbf{k})=\dfrac{ \hbar^2\mathbf{k}^2}{2m} - \mu+\hat{V}(x)-h(\mathbf{k})\hat{\sigma}_3.
\end{equation}
Here,
\begin{equation}
    \hat{V}(x)= (V_0\hat{\sigma}_0-J\mathbf{n}\cdot\hat{\boldsymbol{\sigma}})\delta(x),
\end{equation}
is the spin-dependent interfacial barrier potential and
\begingroup
\allowdisplaybreaks
\begin{align}\label{eqn:h_AM}
     h_\mathbf{k} = &h_0\Bigl[ \sin(2\phi_0)\{\hat{k}_x,\theta(x)\}\hat{k}_y +\cos(2\phi_0)\bigl(\hat{k}_x\theta(x)\hat{k}_x\!-\!\hat{k}_y\theta(x)\hat{k}_y\bigr) \Bigr] \ ,
\end{align}
\endgroup
is the altermagnetic exchange field of an arbitrary orientation. As mentioned in the main text, this term is symmetrized, accounting for the $\theta(x)$ function, which ensures Hermicity across the system. Note that $\hat{k}_i=k_i/k_F$ with $k_F=\sqrt{2m\mu/\hbar^2}$ is a normalized $k$-vector which is dimensionless, allowing us to consider the exchange field $h_0$ as an energy quantity. 

The above Hamiltonian can be diagonalized straightforwardly, yielding the following energy spectrum:
\begin{equation}
    E_\sigma(\vec{k}) = \frac{\hbar^2\vec{k}^2}{2m} - \mu-\zeta_\sigma h_0\qty[2\sin(2\phi_0)\hat{k}_x\hat{k}_y+\cos(2\phi_0)(\hat{k}_x^2-\hat{k}_y^2)],
\end{equation}
where, as before, $\zeta_\sigma=\pm 1$ for $\sigma=\uparrow\!/\!\downarrow$. The above dispersion relations can be easily inverted, arriving at the following $x$ projections (axis of the system) of the wave vectors in the superconductor ($h_0\equiv 0$) and the altermagnet $(h_0\neq0)$, respectively:
\begin{align}
    k_1^\pm \equiv \pm k_1 &= \pm \sqrt{\frac{2m}{\hbar^2}(\mu+E)-k_y^2} \ , \\
    k_\uparrow^\pm\equiv k_2^\pm&=\frac{\pm \mu}{\mu-h_0\cos(2\phi_0)}\sqrt{\frac{2m}{\hbar^2}(\mu+E)\qty[1-\frac{h_0}{\mu}\cos(2\phi_0)]+\qty[\qty(\frac{h_0}{\mu})^2-1]k_y^2}+\nonumber\\
    &+\frac{h_0\sin(2\phi_0)}{\mu-h_0\cos(2\phi_0)}k_y,\\
     k_\downarrow^\pm\equiv k_3^\pm&=\frac{\pm \mu}{\mu+h_0\cos(2\phi_0)}\sqrt{\frac{2m}{\hbar^2}(\mu+E)\qty[1+\frac{h_0}{\mu}\cos(2\phi_0)]+\qty[\qty(\frac{h_0}{\mu})^2-1]k_y^2}-\nonumber\\
     &-\frac{h_0\sin(2\phi_0)}{\mu+h_0\cos(2\phi_0)}k_y.
\end{align}

Here, $\sigma=\uparrow\downarrow$ refers to the spin projection and the superscript $\pm$ refers to the direction of propagation with respect to $k_x>0$. This allows us to identify the incoming and outgoing waves, which are connected by the S-matrix. Note that in our notation the superconductor is characterized by one doubly spin-degenerated channel labeled by 1, and the altergmagnet by two distinct spin channels labeled by 2 and 3, which correspond to the spin-$\uparrow$ band and the spin-$\downarrow$ band, respectively. Since the model used in our work is based on the quasiclassical theory of superconductivity, the S-matrix is constructed by setting $E=0$, i.e., considering only the quasiparticles residing at the Fermi surface. The S-matrix is determined from the boundary conditions imposed on the wave functions at $x=0$ outlined below. The Fermi velocity, which parametrizes the quasiclassical trajectory in the altermagnet, reads
\begin{equation}
    \vec{v}_{F,\sigma} =  \frac{\hbar}{m\mu}\begin{pmatrix}
        [\mu-\zeta_\sigma h_0 \mathrm{cos}(2\phi_0)]k_{F\sigma,x} - \zeta_\sigma h_0 \mathrm{sin}(2\phi_0)k_{F\sigma,y} \\
        [\mu+\zeta_\sigma h_0 \mathrm{cos}(2\phi_0)]k_{F\sigma,y} - \zeta_\sigma h_0 \mathrm{sin}(2\phi_0)k_{F\sigma,x}
    \end{pmatrix},
\end{equation}
Note that in the case of a weak spin polarization, $h_0\ll\mu$, the following holds $\vec{k}_{\uparrow(2)}\approx \vec{k}_{\downarrow(3)} \approx \vec{k}_1$, i.e., the Fermi velocity simply reduces to $\vec{v}_{F,\sigma} \approx (\hbar/m)\vec{k}_F$, which has been used in Sec.~II of the main text.

The boundary conditions imposed on the wave functions at the SC/AM interface are determined following the standard approach based the Schr\"odinger equation
\begin{equation}\label{eqn:SE}
    \hat{H}_\vec{k}\bar{\Psi}_\vec{k}(\vec{r}) = E_\vec{k}\bar{\Psi}_\vec{k}(\vec{r}),
\end{equation}
where $\bar{\Psi}_\vec{k}(\vec{r})=e^{ik_yy}\bar{\psi}(x)$, since we consider a 2D situation, assuming homogeneity in the $y$ direction. Note that the bar symbol $(\bar{\bullet})$ in this context refers to the fact that the wave functions are spinors. Since the transport properties are calculated in the $x$ direction, the first boundary condition is determined simply from the continuity of the wave function throughout the system
\begin{equation}\label{eqn:BC1}
    \bar{\psi}(x-0)=\bar{\psi}(x+0).
\end{equation}
The second boundary condition imposed on the derivative of the wave function at the interface is obtained from the Schr\"odinger equation~\eqref{eqn:SE} itself. We note that the coupling between the momentum dependence of the altermagnetic exchange field and the step function [see Eq.~\eqref{eqn:h_AM}] gives an additional source term, which modifies the standard result obtained for the $\delta$ potential. Specifically, one should take into account
\begin{equation}
    \{\hat{k}_x,\theta(x)\}\!=\!\frac{-i}{k_F}\qty[\partial_x\theta(x)\!+\!\theta(x)\partial_x]\!=\!\frac{-i}{k_F}\qty[\delta(x)\!+\!\theta(x)\partial_x\!+\!\theta(x)\partial_x]\!=\!\frac{-i}{k_F}\qty[\delta(x)\!+\!2\theta(x)\partial_x],
\end{equation}
and
\begin{align}
    k_x\theta(x)k_x\!-\!k_y\theta(x)k_y\!=\!-\frac{1}{k_F^2}\qty[\partial_x\theta(x)\partial_x\!-\!\partial_y\theta(x)\partial_y]\!=\!-\frac{1}{k_F^2}\qty[\delta(x)\partial_x\!+\!\theta(x)(\partial^2_x\!-\!\partial_y^2)],
\end{align}
which results in (for a more detailed derivation see Ref.~\cite{Sun2023})
\begin{align}\label{eqn:BC2}
    \Bigl(1&-\frac{h_0}{{\mu}}\cos(2\phi_0)\hat{\sigma}_3\Bigr)\dv{\bar\psi}{x}\big|_{x=0^+} - \dv{\bar\psi}{x}\big|_{x=0^-}= \frac{2m}{\hbar^2}\qty[V_0\hat{\sigma}_0-J\vec{n}\cdot\bm{\hat{\sigma}}+i\frac{h_0}{{k_F^2}}k_y\sin(2\phi_0)]\bar\psi(x=0).
\end{align}

\begin{figure}[t!]
    \centering
    \includegraphics[width=0.5\linewidth]{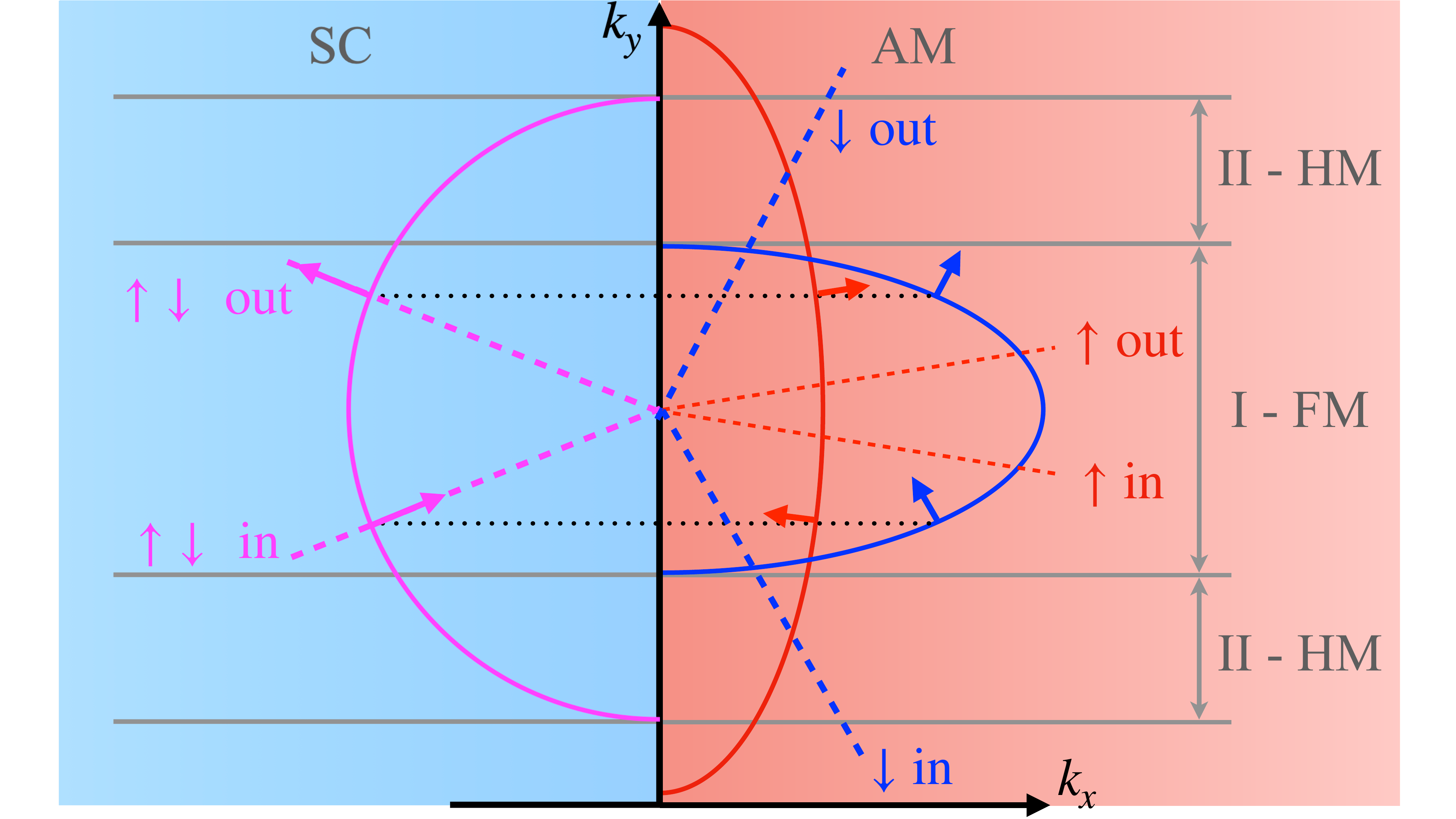}
    \caption{The Fermi surface geometry of a superconductor/altermagnet proximity contact, where latter features the $d_{x^2-y^2}$ orientation ($\phi_0=\pi/4$). For simplicity, we assume the superconducting Fermi surface to be spherical. On both sides the solid arrows perpendicular to the Fermi surface line refer to the corresponding Fermi velocity that parametrizes the quasiclassical trajectory. The latter are denoted by dotted lines: magenta on the superconducting side and red/blue corresponding to spin-$\uparrow$/spin-$\downarrow$ band in the altermagnet. Due to the Fermi surface mismatch and the conservation of the parallel component of the momentum, $k_y$, this altermagnetic orientation may feature a significant half-metallic contribution to transport (see the II-HM regions) }
    \label{fig:Fermi_geometry}
\end{figure}

Equations~\eqref{eqn:BC1} and~\eqref{eqn:BC2} are used for calculating the S-matrix. We do not present the details of the calculation here, as the final expressions are lengthy; however, they are obtained straightforwardly.  We note that, in the general case, the S-matrix has a $4\times4$ matrix structure in combined channel (superconductor/altermagnet)-spin space, as shown in Eq.~(31) in the main text:
\begin{equation}\label{eqn:S_matrix1}
    \bm{S} = \begin{pmatrix}
        \hat{R}_1 & |T_{12}\rangle & |T_{13}\rangle \\
        \langle T_{21}| & r_{22} & r_{23} \\
        \langle T_{31}| & r_{32} & r_{33} 
    \end{pmatrix}.
\end{equation}
The above matrix structure refers to the fact that a superconductor is characterized by one doubly spin-degenerate channel (denoted by 1) and the altermagnet by two spin channels, $\uparrow\!/\!\downarrow$, denoted by 2/3. The S-matrix is calculated under the assumption of a conserved parallel component of the momentum to the interface, $k_y$. This means that, depending on the Fermi surface geometry, for certain values of $k_y$ in the altermagnet, just one spin channel can be available, making it effectively a half-metal. An example of the Fermi surface geometry is depicted in Fig.~\ref{fig:Fermi_geometry}. For illustration, we show the case of the $d_{x^2-y^2}$ orientation of the altermagnet, which does not affect the general approach discussed here. Note that for values {$k_y>\sqrt{2m/(\mu+h_0)}/\hbar$} (note that this is valid for this particular case of the $d_{x^2-y^2}$ orientation), the half-metallic contribution takes place and the S-matrix has an effective $3\times 3$ matrix structure
\begin{equation}\label{eqn:S_matrix2}
    \bm{S} = \begin{pmatrix}
        \hat{R}_1 & T_{12} \\
         T_{21}& r_{22}  
    \end{pmatrix}.
\end{equation}

Finally, we comment on the notation used throughout the paper. In the quasiclassical theory of superconductivity in the ballistic limit, a trajectory of propagation is along the direction of the Fermi velocity, determined by a particular point on the Fermi surface. Figure~\ref{fig:Fermi_geometry} shows one possible scattering event. The dotted lines represent the incoming and outgoing waves (magenta on the SC side, and blue and red on the AM side). The solid arrows perpendicular to the Fermi surface denote the Fermi velocity directions. On the AM side, the blue/red arrow corresponds to the Fermi velocity of the spin-$\uparrow$/spin-$\downarrow$ band, determining the direction of propagation in the altermagnet. Note that the dotted lines are in the directions of the respective arrows, just translated to the common origin, since the Fermi surfaces are elliptical, unlike the Fermi surface of the superconductor, which is assumed to be spherical. The discussed S-matrix is used for solving the effective boundary condition problem following Refs.~\cite{eschrigScatteringProblemNonequilibrium2009,greinSpinDependentCooperPair2009,greinTheorySuperconductorferromagnetPointcontact2010}, as outlined below. 
\section{Boundary conditions}
In solving the boundary condition problem, the central role is played by the renormalized Green's function, which is defined in channel space and incorporates elementary processes that occur at the SC/AM interface:
\begin{equation}
    \bm{\gamma}^\prime\,= \bm{S}\bm{\gamma}\bm{\tilde{S}}\quad\text{and}\quad
    \bm{\tilde\gamma}^\prime\,= \bm{\tilde{S}}\bm{\tilde\gamma}\bm{S},
\end{equation}
where $\bm{S}$ is the S-matrix and $\bm{\gamma}=\gamma_k\delta_{kk'}$ with $k$ denoting the scattering channel. In the case of a strongly spin-polarized altermagnet, the above matrix has the form
\begin{equation}
    \bm{\gamma} = \left(\begin{array}{ccc}
         \hat{\gamma}_1 & {} & {}\\
         & \gamma_2 &  \\
         & & \gamma_3
    \end{array}\right)
\end{equation}
using the same notation as before. The S-matrix in the particle sector is, in general, given by Eq.~\eqref{eqn:S_matrix1}, while the corresponding S-matrix in the hole sector is obtained simply by $\tilde{S}_{ij}=S_{ij}^\ast$. As mentioned in the main text, the $\hat{R}$ block describes the reflection (normal and spin-flip) in the SC, $r_{ij}$ amplitudes correspond to reflections in the AM, and $|T_{ij}\rangle=(t_{ij},t_{ij}^\prime)$ denotes the transmission amplitudes.

Following the notation of Ref.~\cite{eschrigScatteringProblemNonequilibrium2009}, we denote the incoming amplitudes with small letters, $\gamma$ and $\tilde\gamma$, and the outgoing ones with the capital ones, $\Gamma$ and $\tilde\Gamma$. Note that since $\gamma$ and $\tilde\gamma$ have the opposite group (Fermi) velocities, we should also make a distinction between the incoming and outgoing directions, denoted by $p,p^\prime, \dots$ and $k,k^\prime,\dots$, respectively. Considering the \textit{outgoing directions}, the quasiclassical GF is constructed from $\tilde\gamma$ and $\Gamma$, i.e.,
\begin{equation}
    \hat{g}_k=\hat{g}_k[\Gamma,\tilde\gamma].
\end{equation}
To obtain the outgoing $\Gamma$ amplitude, we introduce auxiliary functions such as:
\begin{equation}
    \bm{\mathcal{F}} = \bm{\gamma}^\prime(\bm{1}-\bm{\tilde\gamma}\bm{\gamma}^\prime)^{-1}=\bm{\mathcal{G}}\bm{\gamma}^\prime\quad\text{and}\quad
    \bm{\mathcal{G}} = (\bm{1}-\bm{\gamma}^\prime\bm{\tilde\gamma})=\bm{1}+\bm{\mathcal{F}\bm{\tilde\gamma}}.
\end{equation}
Finally, the outgoing amplitude $\Gamma_k$ is simply calculated from 
\begin{equation}\label{eqn:Gamma_out}
    \Gamma_k = [\mathcal{G}_k]^{-1}\mathcal{F}_k,
\end{equation}
where $\mathcal{G}_k=\mathcal{G}_{kk}$ and $\mathcal{F}_k=\mathcal{F}_{kk}$.
A similar consideration can be applied to incoming directions described by the quasiclassical propagator
\begin{equation}
    \hat{g}_p=\hat{g}_p[\gamma,\tilde\Gamma].
\end{equation}
The corresponding auxiliary functions are given by
\begin{equation}
    \bm{\tilde\mathcal{F}} = \bm{\tilde\gamma}^\prime(\bm{1}-\bm{\gamma}\bm{\tilde\gamma}^\prime)^{-1}=\bm{\tilde\mathcal{G}}\bm{\gamma}^\prime\quad\text{and}\quad
    \bm{\tilde\mathcal{G}} = (\bm{1}-\bm{\tilde\gamma}^\prime\bm{\gamma})=\bm{1}+\bm{\tilde\mathcal{F}\bm{\gamma}},
\end{equation}
leading to the solution for the outgoing amplitude $\tilde{\Gamma}$:
\begin{equation}\label{eqn:GammaTilde_out}
    \tilde\Gamma_p = [\tilde{\mathcal{G}}_p]^{-1}\tilde{\mathcal{F}}_p.
\end{equation}

\section{CPRs for additional altermagnetic orientations}\label{appendix:CPR_different_orientations}

\begin{figure}[t!]
   \centering
    \includegraphics[width=0.7\linewidth]{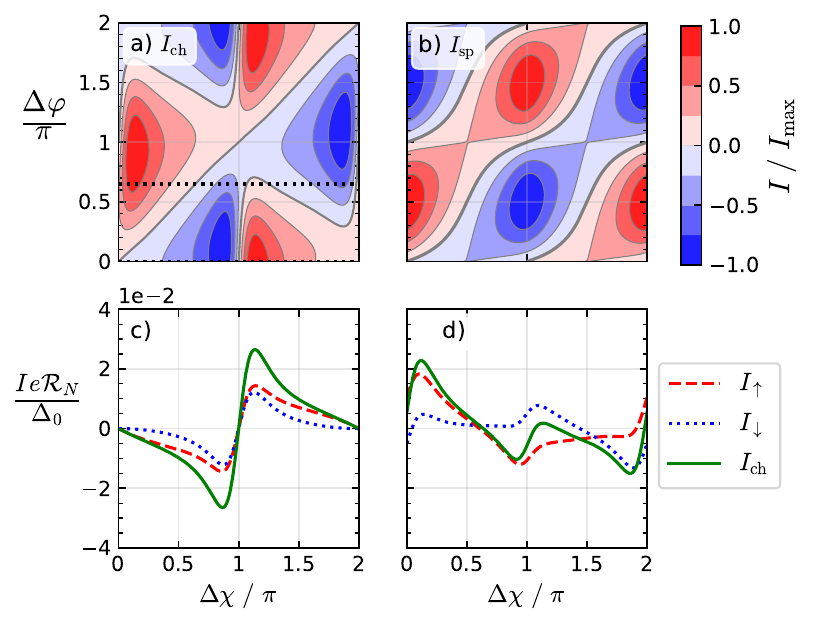}
    \caption{\underline{Altermagnetic orientation $\phi_0=\pi/8$}. Panel (a): the normalized charge current $I_\mathrm{ch}$ as a function of the superconducting phase difference $\Delta\chi$ and the quantum-geometric phase $\Delta\varphi$. The spin-resolved and total charge current-phase relations for calculated for $\Delta \varphi=0$ and $\Delta\varphi=0.65\pi$ are shown in panels (c) and (d), respectively. The system's parameters are as in Fig.~4 of the main text. 
    }
    \label{fig:CPR_0125_orientation}
\end{figure}

In this section, we present the Josephson CPRs for two more altermagnetic orientations, to support the conclusions from the main text [see Sec.~IIIB]. Figure~\ref{fig:CPR_0125_orientation} shows the case of the $\phi_0=\pi/8$ orientation of the altermagnet. Panel (a) shows the charge current $I_\ch$ as a function of the superconducting phase difference $\Delta\chi$ and the quantum geometric phase $\Delta\varphi$ calculated for the same parameters as in Fig.~4 of the main text. Panel (b) shows the corresponding spin current. Panel (c) displays the spin-resolved CPRs (dashed and dotted lines) and the total charge CPR (green solid line) obtained for $\Delta\varphi=0$. Since the system features a coplanar magnetization profile, the normal Josephson effect emerges. In contrast, panel (d) shows the corresponding CPRs obtained for $\Delta\varphi=0.65\pi$ [see the dotted line in pane (a)]

\begin{figure}[t!]
    \centering
    \includegraphics[width=0.7\linewidth]{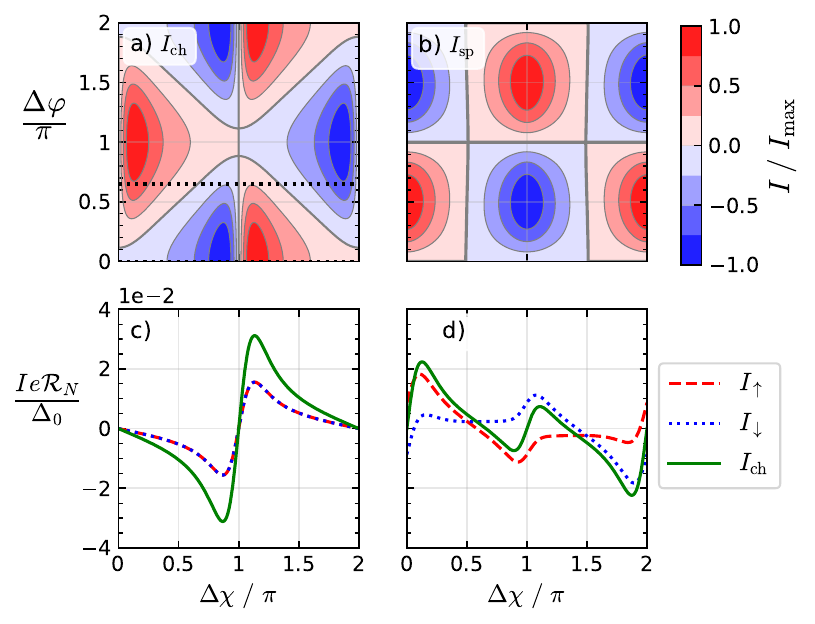}
    \caption{Analogous figure to Fig.~\ref{fig:CPR_0125_orientation} obtained for the the $d_{xy}$ orientation of the altermagnet, i.e., $\phi_0=\pi/4$.
    }
    \label{fig:CPR_025_orientation}
\end{figure}

Figure~\ref{fig:CPR_025_orientation} shows fully analogous results as that of Fig.~\ref{fig:CPR_0125_orientation} obtained for the $d_{xy}$ orientation of the altermagnet characterized by $\phi_0=\pi/4$. Interestingly, this altermagnetic orientation yields no Josephson diode effect. The explanation lies in the fact that the Josephson diode effect in strongly spin-polarized magnetic systems vital depends on the presence of the half-metallic contribution to the CPR~\cite{greinSpinDependentCooperPair2009,schulz2025_prl,schulz2025_prb,nikolic_2025}, which is absent in the case of the $d_{xy}$ orientation of the altermagnet. In contrast, the $d_{x^2-y^2}$ orientation whose Fermi surface geometry is shown in Fig.~\ref{fig:Fermi_geometry} maximizes the half-metallic contribution maximizing the Josephson diode effect.

\newpage
\newpage

\bibliography{references}